\documentclass[aps,pra,superscriptaddress,floatfix,onecolumn,floatfix,notitlepage,nofootinbib]{revtex4-1}
\usepackage{amsfonts,color,bigints}
\usepackage{amsmath}
\usepackage{amssymb}
\usepackage{physics}
\usepackage{graphicx,dsfont}
\usepackage[utf8]{inputenc}
\usepackage[colorlinks=true,citecolor=blue,linkcolor=red]{hyperref}
\usepackage{newtxtext,newtxmath}
\DeclareMathAlphabet{\mathcal}{OMS}{cmsy}{m}{n}

\newcommand{\spin}{\hat{J}}

\newcommand{\qham}{\hat{\mathcal{H}}}
\newcommand{\beq}{\begin{equation}}
\newcommand{\eeq}{\end{equation}}
\newcommand{\C}{\hat{\mathcal{C}}}
\newcommand{\K}{\hat{\mathcal{K}}}
\newcommand{\eps}{\varepsilon}

\renewcommand\labelenumi{(\roman{enumi})}
\renewcommand\theenumi\labelenumi
\newcommand{\hatmath}[1]{\hat{\mathcal{#1}}} 

\begin{document}

\title{Relaxation time as a control parameter for exploring  dynamical phase diagrams}

\author{\'{A}ngel L. Corps}
    \email[]{corps.angel.l@gmail.com}
    \affiliation{Instituto de Estructura de la Materia, IEM-CSIC, Serrano 123, E-28006 Madrid, Spain}
    \affiliation{Grupo Interdisciplinar de Sistemas Complejos (GISC),
Universidad Complutense de Madrid, Av. Complutense s/n, E-28040 Madrid, Spain}

\author{Pedro P\'{e}rez-Fern\'{a}ndez}
    \email[]{pedropf@us.es}
    \affiliation{Departamento de F\'{i}sica Aplicada III, Escuela T\'{e}cnica Superior de Ingenier\'{i}a, Universidad de Sevilla, E-41092 Sevilla, Spain}
    \affiliation{Instituto Carlos I de F\'{i}sica Te\'{o}rica y Computacional, Universidad de Granada, Fuentenueva s/n, E-18071 Granada, Spain}
    
\author{Armando Rela\~{n}o}
    \email[]{armando.relano@fis.ucm.es}
    \affiliation{Grupo Interdisciplinar de Sistemas Complejos (GISC),
Universidad Complutense de Madrid, Av. Complutense s/n, E-28040 Madrid, Spain}
    \affiliation{Departamento de Estructura de la Materia, F\'{i}sica T\'{e}rmica y Electr\'{o}nica, Universidad Complutense de Madrid, Av. Complutense s/n, E-28040 Madrid, Spain}

\begin{abstract}
We explore a full dynamical phase diagram by means of a double quench protocol that depends on a relaxation time as the only control parameter. The protocol comprises two fixed quenches and an intermediate relaxation time that determines the phase in which the quantum state is placed after the final quench. We apply it to an anharmonic Lipkin-Meshkov-Glick model. We show that its two excited-state quantum phase transitions split the spectrum into three different phases: two symmetry-breaking ones, characterized by different constants of motion, and a disordered phase. As a consequence, our protocol allows us to explore all the dynamical phase transitions arising from two kind of quenches: the typical one, leading the system from a symmetry-breaking to a disordered phase, and another one in which the system transites between two different symmetry-breaking phases. We characterize all of them in terms of the constants of motion appearing in all three phases of the model.

\end{abstract}
\date{\today}

\maketitle

\section{Introduction}

The study of phase transitions and critical phenomena has attracted much attention in recent years. They appear in a wide range of physical settings, including nuclear physics, molecular physics, quantum optics and condensed matter. Furthermore, the recent development of quantum technologies has placed them right at the core  of fascinating effects \cite{Yuzbashyan2006,Baumann2010,Gring2012,Chu2020, Muniz2020,MunozArias2023}.

Besides thermal phase transitions, driven by temperature, quantum systems display several kinds of critical phenomena driven by non-thermal parameters. A quantum phase transition (QPT) entails a non-analytical evolution of the ground state of the system as a function of some control parameters, and impacts on the ground state wavefunction and different observables \cite{Carr,Sachdev1999}. A similar kind of phase transition may also occur in excited states, giving rise to excited state quantum phase transitions (ESQPTs) \cite{Cejnar2021}. Their study is an active research field with a deep impact on different branches of physics. Since they were originally proposed in \cite{Caprio2008}, ESQPTs have been shown to have important dynamical consequences in many-body quantum systems in nuclear and molecular physics \cite{Caprio2008, Dong2021,Macek2019,Khalouf2021,Khalouf2022}, quantum optics \cite{Perez2011, Perez2011b, Brandes2013, Puebla2013, Puebla2016, Bastarrachea2014,Wang2020, Chavez2022,Corps2022JPA,Corps2022PRA},  and condensed matter physics \cite{Bastidas2014, Bastidas2014b, Wang2021, Mondal2022,Relano2008, Perez2009, GamitoarXiv, KhaloufarXiv, Nader2021,Tian2020, Feldmann2021, Cabedo2021, Meyer2023, Zhou2022,Ribeiro2008}. Just to mention a few of those consequences, ESQPTs enhance the decoherence process in a system \cite{Relano2008,Perez2009}, affect processes of quench dynamics \cite{Perez2011, Santos2015, Lobez2016, PBernal2017, Kloc2018, Wang2023} and quantum work statistics \cite{Kopylov2015}, and localization \cite{Wang2017}. Also, ESQPTs show universal dynamical scaling \cite{Puebla2020}, symmetry-breaking equilibrium states \cite{Puebla2013, Puebla2015}, dynamical instabilities \cite{Bastidas2014, Bastidas2014b}, irreversibility without energy dissipation \cite{Puebla2015} and reversible quantum information spreading \cite{Hummel2019}; they can be related with thermal phase transitions \cite{Perez2017} and dynamical phase transitions (DPTs) \cite{Puebla2020, Corps2022PRB, Corps2022arXiv,Corps2023arXiv,Corps2023PRB}. For a detailed review on ESQPTs and its physical manifestations, the interested reader may consult \cite{Cejnar2021}. 

The last phenomenon, DPTs, involves two different kinds of phase transitions. DPTs of the first type, DPTs-I, are linked to the oscillations around a kind of stationary value observed in the expectation value of certain observables \cite{Marino2022}. They are usually triggered by a quantum quench, that is, a sudden change in a control parameter in the Hamiltonian of the system, which takes it out of equilibrium. A DPT-I occurs when, for a particular critical quench, the expectation value of an observable changes from fluctuating around zero to fluctuating around a certain finite value. Hence, they can be characterized by means of the long-time average of such observable \cite{Muniz2020,Chu2020,Eckstein2008,Moeckel2008,Eckstein2009,Sciolla2011,Smale2019,Lang2018concurrence,Lerose2019, LewisSwan2021}. DPTs of the second type, DPTs-II, occur when certain return probabilities become non-analytical at particular critical times \cite{Heyl2013,Heyl2014,Heyl2018,Heyl2019,Tian2020,Halimeh2017prethermalization,Mori2018,Jurcevic2017,Sciolla2013,Mishra2020,Halimeh2021,Halimeh2021ES,Halimeh2017,Zauner2017,Lang2018,Hashizume2022,Karrasch2013, Bhattacharya2017,Zunkovic2018,Vajna2014, Sehrawat2021,Zhou2019,Naji2022}. Usually, they also appear as a consequence of a quench, but they are not directly linked to long-time averages or to equilibrium values.

Even though DPTs and ESQPTs refer to fundamentally different physical effects, it has been recently shown that under some conditions both can be triggered by the appearance of certain conserved charges, at least in systems with long or infinite-range interactions \cite{Corps2022PRB,Corps2022arXiv,Corps2023arXiv}. In this paper, we generalize such results to systems having two different symmetry-breaking phases, and we devise a protocol to explore the corresponding full phase diagram, which includes two different ESQPTs and both kinds of DPTs. The protocol consists of three steps. First, we prepare the system in a symmetry-breaking ground state (or a symmetry-breaking highest excited state, which is the ground state of the same Hamiltonian with a negative global sign). Then, we perform a first quench to lead the system onto a symmetry-breaking excited-state phase of the same nature. And finally, after a controlled relaxation time in this intermediate stage, we do a second quench. The trademark of this protocol is that its only changing variable is the intermediate relaxation time. Hence, just by changing an easily tunable parameter, we can explore three different phases: a symmetry-breaking phase of the same nature of the initial one, a symmetry-breaking phase of a different nature, and a disordered one. As a consequence, we can study the DPTs arising from two different kinds of quenches: one between two different symmetry-breaking phases, and another one between a symmetry-breaking and a disordered one.

We apply this protocol to the anharmonic Lipkin-Meshkov-Glick model (aLMG). The main feature of this model is that it undergoes two different ESQPTs, some of whose static and dynamical properties are studied in \cite{GamitoarXiv, KhaloufarXiv}; thus, it is the perfect choice for the purpose we pursue in this work. To begin with, we rely on the tools developed in \cite{Corps2021,Corps2022PRB,Corps2022arXiv} to show that we can characterize each of its two symmetry-breaking phases by its own pair of non-commuting constants of motion, and hence to conclude that they are of a different nature. Then, we extend the generalized microcanonical ensemble (GME) proposed in \cite{Corps2022PRB,Corps2022arXiv} in order to properly describe the long-time average of relevant observables, taking into account all the constants of motion required to describe each of the spectral phases. Finally, as our protocol allows us to explore the DPTs arising from two different kinds of quenches, we unveil the role played by the constants of motion in all the DPTs-I and DPTs-II arising from them.

The paper is organized as follows: in Sec. \ref{sec1} we introduce the model to which the double quench is applied; in particular, Sec. \ref{semiclassical} reviews its classical limit.  
In Sec. \ref{constants} we define four constants of motion that can be used to identify the three different dynamical phases emerging from the two ESQPTS. Also, in Sec. \ref{GME}, we extend the generalized microcanonical ensemble (GME) introduced in \cite{Corps2022arXiv}. Sec. \ref{protocol} presents a double-quench protocol that allows us to explore the full phase diagram,
which is crucial to connect phases of the same nature, i.e. a symmetry-breaking phase with another one. Sec. \ref{DPT-I} is devoted to DPTs-I, where we focus on the evolution and the long-time average of spin operators. In Sec. \ref{DPT-II} presents our findings on DPTs-II; here, we compare the parity-projected return probability against the standard survival probability. Finally, we conclude in Sec. \ref{conclusions}.

\section{Model Hamiltonian} \label{sec1}

We start from a Heisenberg $XYZ$ model of $N$ spin-$1/2$ particles with an infinite-range interaction. The system Hamiltonian is given by
\begin{equation}
\hat{\mathcal{H}}= h \sum_{i=1}^{N} \hat{\sigma}_{i}^{z} + h_{x}\sum_{i<j=1}^{N}\hat{\sigma}_{i}^{x}\hat{\sigma}_{j}^{x}+h_{z}\sum_{i<j=1}^{N}\hat{\sigma}_{i}^{z}\hat{\sigma}_{j}^{z}~,
\label{eq:LMG0}
\end{equation}
where $\hat{\sigma}_{i}^{\beta}$ are the Pauli spin-$1/2$ matrices acting on site $i = 1, 2, \ldots, N$, and $\beta = x,y,z$. Here, we assume $\hbar=1$, $h$ is the magnitude of an external magnetic field, and $h_{x}$ and $h_{z}$ are, respectively, the interaction strength along the $x-$ and $z-$axes. Introducing collective spin operators $\spin_{\beta} = \frac{1}{2} 
\sum_{i=1}^{N}\hat{\sigma}_{i}^{\beta}$ for $\beta = x,y,z$, adding a constant term $j(1-h_{x}j)+2h_{z}j(j+1)$ and identifying the terms $h_{x}J=-\xi$, 
$h=\frac{1}{2}(1-\xi+\frac{\alpha}{2j}(2j+1))$ and $h_{z}=\frac{\alpha}{4j}$, it is possible to map this Hamiltonian to an anharmonic Lipkin-Meshkov-Glick model (aLMG) in terms of only two control parameters $\xi \in [0,1]$ and $\alpha \in \mathbb{R}$,
\begin{equation}
\hat{\mathcal{H}} = (1-\xi)\left(j + \spin_{z} \right)
+\frac{2\xi}{J}\left(j^2-\spin_{x}^2\right)
+\frac{\alpha}{2j}\left(j+\spin_z\right)\left(j+\spin_z+1\right).
\label{eq:LMG2}
\end{equation}
When 
$\alpha=0$ we recover the well-known Lipkin-Meshkov-Glick model, while $\alpha \neq 0$ introduces an anharmonic perturbation.
In all cases, the total collective spin operator $\mathbf{\spin}^2$ is conserved, $[\hat{H}, \mathbf{\spin}^2]=0$. This allows us to split the Hamiltonian matrix in different blocks corresponding to different irreducible representations, labeled with the eigenvalues of $ \mathbf{\spin}^2$, $j(j+1)$. Since the ground state is included in the maximum irreducible representation of the system, we focus on the sector with $j=N/2$. Furthermore, this allows us to reduce the dimension of the Hilbert space from exponentially, $2^N$, to only linearly large in the number of sites, $N+1$.

Irrespective of the value of $\alpha$, the Hamiltonian \eqref{eq:LMG2} also conserves a discrete symmetry, $\hat{\Pi}=e^{i\pi(j+\spin_z)}$, which is the parity operator. This operator divides the spectrum in two blocks, one with even parity and another one with odd parity. 
Specifically, if $\ket{E_{n,\pm}}$ is the eigenstate corresponding to the eigenvalue $E_{n,\pm}$ with either parity $+$ (even) or $-$ (odd), then $\hat{\Pi}\ket{E_{n,\pm}}=\pm \ket{E_{n,\pm}}$.

\begin{figure}
\begin{centering}
\includegraphics[scale=0.5]{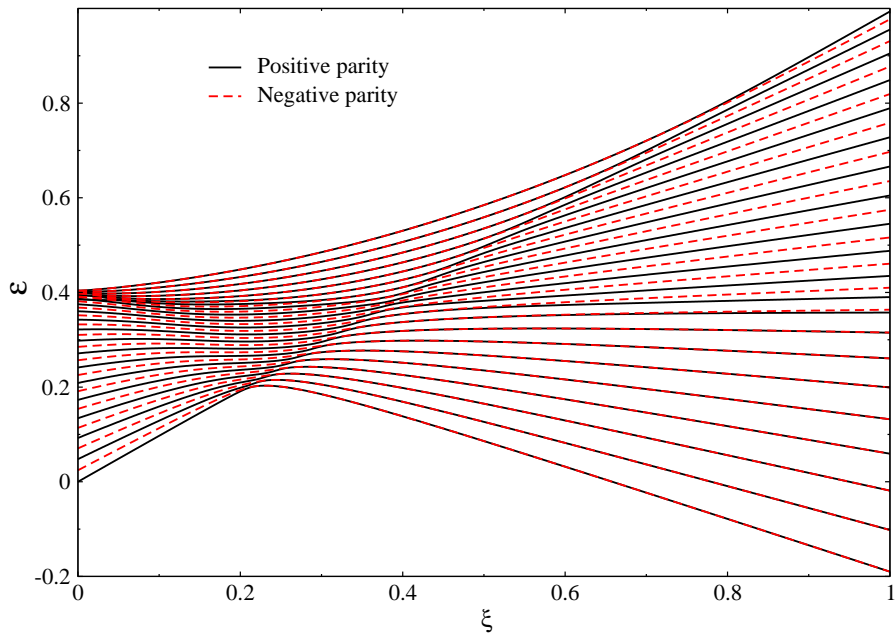}
\par\end{centering}
\caption{Scaled energy spectrum of \eqref{eq:LMG2} for a system size $N=40$ ($j=20$) as a function of the control 
parameter $\xi$ and for a value $\alpha=-0.6$. Solid black lines stand for states with positive parity 
while dashed red lines stand for states with negative parity.}
\label{spectrum}
\end{figure}

In Fig.~\ref{spectrum}, we depict the scaled energy spectrum of the system as a function of the control parameter $\xi$, with $N=40$ particles (or collective spin length $j=20$), and 
$\alpha=-0.6$. The energy scale of this level-flow diagram is given by $\varepsilon\equiv E/N$, where $E$ represents the actual Hamiltonian eigenvalue, which is an extensive quantity. In order to compare with the infinite-size limit of the model, the intensive energy $\varepsilon$ is used when appropriate in this work. The Hamiltonian eigenstates are still denoted as $\ket{E_{n,\pm}}$. There is a QPT for $\xi=0.2$, and there are also two critical lines in excited states 
corresponding with ESQPTs. These ESQPTs are noticeable due to the clustering of energy levels in the 
spectrum. Some of their properties, including some dynamical consequences, have been already studied 
in previous works \cite{GamitoarXiv, KhaloufarXiv}; yet, their characterization in terms of constants of motion has not been discussed. For our purposes, the main trademark if this model is that, for $\xi > 0.2$, we observe two different symmetry-breaking phases, characterized by the presence of degenerate parity doublets, and a disordered phase, where these degeneracies are broken, in between.

\subsection{Classical limit} \label{semiclassical}

The thermodynamic limit of this model can be explored by means of spin coherent states, $\ket{\omega}$, in the limit $j \rightarrow \infty$ \cite{Radcliffe71}, where $\omega$ is a complex parameter that includes the classical canonical momentum $P$ and position $Q$ variables. The result is a classical Hamiltonian, $H(Q,P)$,
\begin{equation} \label{enfun} H(Q,P) = \frac{\bra{\omega}    
\hat{\mathcal{H}} \ket{\omega}}{N}=\frac{1-\xi}{2}\left(P^2+Q^2\right)+\frac{\alpha}{4}
\left(P^2+Q^2\right)^2+\xi Q^2\left(P^2+Q^2-2\right) + \xi~.
\end{equation}

\noindent It is worth noting that the canonical variables $Q$ and $P$ are constrained as $0\leq Q^2+P^2\leq 2$, so the classical phase space is bounded. Note also that in the classical limit the classical variables $(Q,P)$ are continuous and commute. The number of effective degrees of freedom of the system is $f=1$. The energy scale associated to this intensive classical Hamiltonian is given by the normalized energy $\varepsilon$.

The main properties of this model were analyzed in \cite{GamitoarXiv}. Specifically, the quantum model \eqref{eq:LMG2} exhibits certain phase transitions that are similarly found in the classical limit \eqref{enfun}. For our purposes, these features can be summarized as follows:

(i) A QPT separates a disordered from a symmetry-breaking phase at zero temperature. 

(ii) A similar phenomenon occurs at the highest excited state. It can be understood as a QPT for the same Hamiltonian with a global negative sign.

(iii) There are one or two critical lines, depending on the values of $\alpha$ and $\xi$, giving rise to different ESQPTs.

\begin{figure}
\begin{centering}
\includegraphics[scale=0.8]{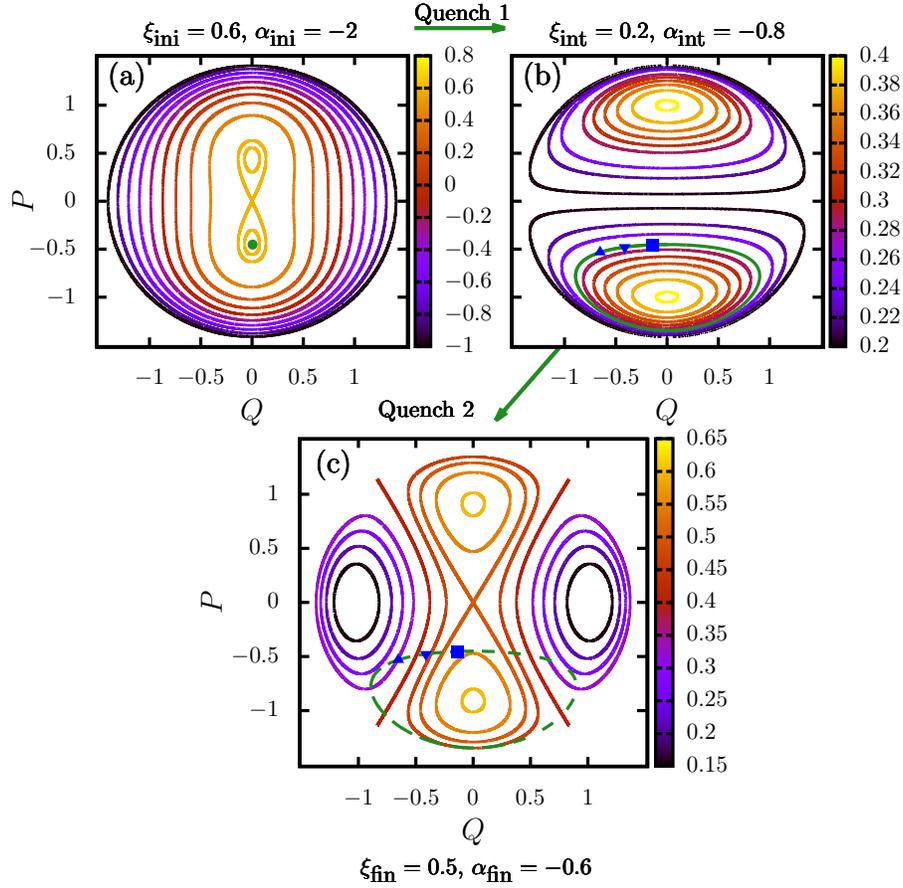}
\par\end{centering}
\caption{Classical orbits in the phase space of the Hamiltonian \eqref{enfun} for different values of the parameters $\xi$ and $\alpha$. The chosen values match those used in the quench protocol detailed in Sec.~\ref{protocol}, and the color of the orbits is determined by its energy (color bars). Panel (a) shows that the phase space exhibits two degenerate maxima and a saddle point. During the quench protocol, we prepare the initial state (green point) in the higher classical energy well and perform the first quench. Panel (b) depicts a disconnected phase space. Each disconnected piece has its own maximum. We let evolve the system during a time $\tau_{\textrm{int}}$ under this Hamiltonian. The markers show the position of the classical state after the quench at different values of time: $\tau_{\textrm{int}}=0.5$ (blue square), $\tau_{\textrm{int}}=1.5$ (downward triangle) and $\tau_{\textrm{int}}=2.5$ (upward triangle). The second quench is performed and the phase space of the final Hamiltonian is depicted in panel (c). There are three distinct regions not connected by any classical orbit. }
\label{fig:contours}
\end{figure}

To illustrate all of these phenomena, we depict the classical orbits in the phase space of the Hamiltonian Eq. 
\eqref{enfun} for different values of the parameters $\xi$ and $\alpha$ in Fig.~\ref{fig:contours}. Each line stands for the classical orbit $(Q,P)$ where the Hamiltonian remains constant at a certain energy value $\varepsilon$, that is, $H(Q,P)=\varepsilon$.  In panel (a), we can observe that there are two degenerate maxima at $Q=0$ but with different signs of $P$. For lower values of the energy, we find pairs of disjoint orbits with the same energy. The value $Q=P=0$ is a saddle point, within a critical orbit that crosses itself. Below this energy, all the orbits are connected, and the lower value in the classical energy is found at the border of the phase space. In panel (b) the phase space is disjoint in two regions; we can see one maximum in each of them. The classical orbits are always confined in one of these regions, with no possibility of crossing at any energy. The maxima are degenerate and occur at $Q=0$, with different signs of $P$, as in the case displayed in panel (a). In panel (c), the phase space is divided into three different regions. We find two degenerate minima in the leftmost and rightmost parts of the phase space; they are degenerate with $P=0$ and different signs of $Q$. For larger energy values, we find disjoint orbits. If energy is further increased, we arrive to a critical and connected orbit, reaching the border of the phase space. Then, as we further increase the energy, the next orbits become connected. Finally, we find a second critical orbit with a saddle point, at $Q=P=0$. From it to the maximum classical energy, the orbits are disjoint again, and we find two degenerate maxima with $Q=0$ and different signs of $P$.

In addition, it is straightforward to compute the expectation value of the operators $\spin_x$, $\spin_y$, and $\spin_z$ using the spin coherent states:
\begin{eqnarray}\label{expected_values}
  \bra{\omega} \spin_x \ket{\omega} &= 2j\frac{\textrm{Re}(\omega)}{|\omega|^2+1} =& jQ\sqrt{2-P^2-Q^2}~, 
\nonumber\\
  \bra{\omega} \spin_y \ket{\omega} &= -2j\frac{\Im(\omega)}{|\omega|^2+1} =& -jP\sqrt{2-P^2-Q^2}~, \\
  \bra{\omega} \spin_z \ket{\omega} &= j\frac{|\omega|^2-1}{|\omega|^2+1} =& j(P^2+Q^2-1)~.\nonumber
\end{eqnarray}
Their dependence with respect to the canonical variables $Q$ and $P$ is very useful to derive a set of constants of motion characterizing all the three spectral phases, as we will see in next section.

\section{Description of dynamical phases in terms of constants of motion} 

Recently, it was shown in \cite{Corps2021}, in a model with just one ESQPT, that the phases in which the ESQPT splits the spectrum can be identified by means of an operator $\C$. This operator is a constant of motion and acts as a discrete symmetry in one of these phases; indeed, the spectrum of $\C$ is simply $\textrm{Spec}(\C)=\{\pm 1\}$. Its origin lays in the disjoint topology of the classical phase space below the critical energy of such ESQPT. 
In the present work, we are interested in a generalization to cover the more complex structure of the classical phase space of the aLMG.

\subsection{Constants of motion characterizing two different symmetry-breaking phases}
\label{constants}

We start by considering that we can distinguish two different disjoint regions and a connected one in its classical phase space, as it is shown in Fig. \ref{fig:contours}(c). We can see that 
this model displays two ESQPTs, at energies $\varepsilon_{c_{1}}$ and $\varepsilon_{c_{2}}$, with two symmetry-breaking phases and two different double-well structures. If $\varepsilon < \varepsilon_{c_{1}}$, trajectories are trapped with either $Q>0$ or $Q<0$. If $\varepsilon_{c_{1}} < \varepsilon < \varepsilon_{c_{2}}$, trajectories explore the whole available phase space. And if $\varepsilon > \varepsilon_{c_{2}}$ trajectories are trapped again, with either $P>0$ or $P<0$. Thus, following the spirit of \cite{Corps2021} and taking into account Eq. \eqref{expected_values}, it seems logical to rely on $\hat{\mathcal{C}}_x = \textrm{sign} (\hat{J}_x)$ to identify the first ESQPT, and on $\hat{\mathcal{C}}_y = \textrm{sign} (\hat{J}_y)$ as a signature of the second one. As pointed out in \cite{Corps2022arXiv,Corps2022PRB}, in the case of $\mathbb{Z}_{2}$ symmetry-broken systems the existence of $\C$ immediately implies that there is a third constant of motion, $\hat{\mathcal{K}} = (i/2) [\hat{\mathcal{C}}, \hat{\Pi}]$, closing a SU(2) structure.

\begin{figure}[h!]
\begin{centering}
\includegraphics[scale=0.8]{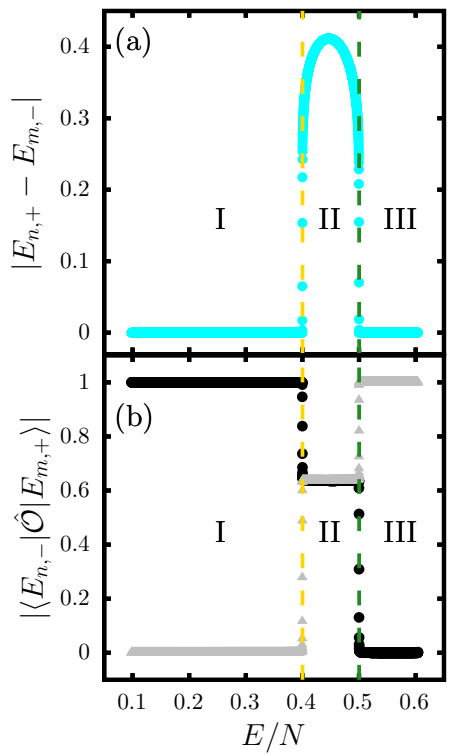}
\par\end{centering}
\caption{(a) Energy gap of states of closest eigenvalues of opposite parity. (b) Expectation value of $\hat{\mathcal{C}}_{x}$ (black) and $\hat{\mathcal{C}}_{y}$ (gray) taken in the eigenstates with eigenvalues of opposite parity whose gap is in (a). The Hamiltonian is Eq. \eqref{eq:LMG2} with $\xi=0.5$ and $\alpha=-0.6$. System size is $j=3200$.  }
\label{fig:paneldeg}
\end{figure}

To test if this idea is correct, we focus on the Hamiltonian eigenvalues. 
In Fig. \ref{fig:paneldeg}(a) we have represented the energy gap of the two states of the Hamiltonian \eqref{eq:LMG2} of opposite parity that are closest in energy, $|E_{n,+}-E_{m,-}|$ ($n=m$ in phase I but $n\neq m$ in phase III), as a function of energy. Because we will be interested in the dynamics of the final stage of our protocol with $\xi=0.5$ and $\alpha=-0.6$, this is the case that we examine here. It is clearly observed that in phases I and III the gap is vanishingly small (it vanishes exponentially with $N$), while in phase II there are no such degeneracies. The boundaries of these phases are clearly demarcated by the ESQPTs at energies $\epsilon_{c_1}$ and $\epsilon_{c_2}$. Thus, phases I and III are symmetry-broken phases, while phase II is not. 

In Fig. \ref{fig:paneldeg}(b) we focus on $\bra{E_{n,-}} \hat{\mathcal{C}}_x \ket{E_{m,+}}$ and $\bra{E_{n,-}} \hat{\mathcal{C}}_y \ket{E_{m,+}}$, where $\ket{E_{n,-}}$ and $\ket{E_{m,+}}$ represent a pair of (almost) degenerate eigenstates. The key feature is that $\left| \bra{E_{n,-}} \hat{\mathcal{C}}_x \ket{E_{m,+}} \right|$ is one in phase I and zero in phase III, whereas $\left| \bra{E_{n,-}} \hat{\mathcal{C}}_y \ket{E_{m,+}} \right|$ behaves exactly in the opposite way. It is also worth noting that neither $\left| \bra{E_{n,-}} \hat{\mathcal{C}}_x \ket{E_{m,+}} \right|$ nor $\left| \bra{E_{n,-}} \hat{\mathcal{C}}_y \ket{E_{m,+}} \right|$ is one in phase II.

To understand the relevance of these results, let us consider an arbitrary initial state $\ket{\Psi(0)}$ evolving in time as $\ket{\Psi(t)}=e^{-i\hatmath{H}t}\ket{\Psi(0)}=\sum_{n,k}c_{n,k}e^{-iE_{n,k}t}\ket{E_{n,k}}$, with $n\in\mathbb{N}$ and $k\in\{+,-\}$. The instantaneous expectation value of a physical observable $\hatmath{O}$ at time $t$, $\langle\hatmath{O}(t)\rangle\equiv \bra{\Psi(t)}\hatmath{O}\ket{\Psi(t)}$, is  
\begin{equation}\label{eq:Ot}
    \langle\hatmath{O}(t)\rangle=\sum_{n,k}\sum_{m,\ell}c_{n,k}c_{m,\ell}^{*}e^{-i(E_{n,k}-E_{m,\ell})t}\bra{E_{m,\ell}}\hatmath{O}\ket{E_{n,k}}.
\end{equation}
Although, mathematically speaking, such an instantaneous value remains oscillating at all times, one may consider a time asymptotic value around which this instantaneous value fluctuates for sufficiently long times. This is the long-time average given by
\beq\label{eq:longtime}
\overline{\langle \hatmath{O}\rangle}\equiv \lim_{\tau\to\infty}\frac{1}{\tau}\int_{0}^{\tau}\textrm{d}t\,\langle\hatmath{O}(t)\rangle.
\eeq
Such asymptotic value is commonly used to define order parameters of dynamical phases. From Eq. \eqref{eq:Ot}, a dependence on the spectrum properties is immediately expected. In phase II, where there are no degeneracies, the oscillating off-diagonal terms $\propto e^{-i(E_{n,+}-E_{n,-})t}$ eventually average out to zero, and the only contribution comes from diagonal terms. This yields the general expression
\beq
\overline{\langle \hatmath{O}\rangle}_{\textrm{II}}=\sum_{n}[|c_{n,+}|^{2}\bra{E_{n,+}}\hatmath{O}\ket{E_{n,+}}+|c_{n,-}|^{2}\bra{E_{n,-}}\hatmath{O}\ket{E_{n,-}}],
\eeq
valid for any $\hatmath{O}$. However, in phases I and III, there are degeneracies (assume that these occur for the pairs $E_{n,+}$ and $E_{n,-}$ for notational simplicity), which means that off-diagonal terms also survive the time average in Eq. \eqref{eq:longtime}: 
\beq
\overline{\langle \hatmath{O}\rangle}_{\textrm{I,III}}=\sum_{n}[|c_{n,+}|^{2}\bra{E_{n,+}}\hatmath{O}\ket{E_{n,+}}+|c_{n,-}|^{2}\bra{E_{n,-}}\hatmath{O}\ket{E_{n,-}}]+\sum_{n}[c_{n,+}c_{n,-}^{*}\bra{E_{n,-}}\hatmath{O}\ket{E_{n,+}}+c_{n,-}c_{n,+}^{*}\bra{E_{n,+}}\hatmath{O}\ket{E_{n,-}}].
\eeq

Let us focus first in the disordered phase, that is, in phase II. 
Because the sign operators $\C_{x,y}$ and $\K_{x,y}$ only connect eigenstates of opposite parity, the only non-zero matrix elements entering Eq. \eqref{eq:Ot} are the time-dependent non-diagonal ones, whereas the diagonal expectation values vanish identically. Thus, we immediately obtain two important conclusions: (i) none of the operators $\hat{\mathcal{C}}_{x}$, $\hat{\mathcal{C}}_{y}$, $\hat{\mathcal{K}}_{x}$ and $\hat{\mathcal{K}}_{y}$ is a constant of motion within phase II, and (ii) all of their expectation values vanish in this phase
\beq
\overline{\langle \C_{x,y}\rangle}_{\textrm{II}}=0,\,\,\,\overline{\langle \K_{x,y}\rangle}_{\textrm{II}}=0. 
\eeq
Note that these equalities hold irrespective of the coefficients $c_{n,\pm}$, which means that they are satisfied by \textit{any} initial state $\ket{\Psi(0)}$. That is, none of these operators provides relevant information within phase II.

For phases I and III, the contribution of the non-diagonal terms in Eq. \eqref{eq:longtime} can be inferred from Fig. \ref{fig:paneldeg}(b). As $\left| \bra{E_{n,-}} \hat{\mathcal{C}}_x \ket{E_{n,+}} \right|=\left| \bra{E_{n,-}} \hat{\mathcal{K}}_x \ket{E_{n,+}} \right|=1$ in phase I, the only non-zero matrix elements in Eq. \eqref{eq:Ot} are precisely the non-diagonal elements appearing in Eq. \eqref{eq:longtime}. On the contrary, as $\left| \bra{E_{n,-}} \hat{\mathcal{C}}_y \ket{E_{n,+}} \right|=\left| \bra{E_{n,-}} \hat{\mathcal{K}}_y \ket{E_{n,+}} \right|=0$ the corresponding non-diagonal elements in Eq. \eqref{eq:longtime} are identically zero, and therefore all the non-zero matrix elements in Eq. \eqref{eq:Ot} are time-dependent. Therefore, we conclude that only $\hat{\mathcal{C}}_x$ and $\hat{\mathcal{K}}_x$ are constant in phase I, and that
\beq
\overline{\langle \C_{x}\rangle}_{\textrm{I}}=2\sum_{n}\textrm{Re}\left(c_{n,+}c_{n,-}^{*}\bra{E_{n,-}}\C_{x}\ket{E_{n,+}}\right),\,\,\,\overline{\langle\K_{x}\rangle}_{\textrm{I}}=2\sum_{n}\textrm{Im}\left(c_{n,+}c_{n,-}^{*}\bra{E_{n,-}}\K_{x}\ket{E_{n,+}}\right),\,\,\,\overline{\langle\C_{y}\rangle}_{\textrm{I}}=0,\,\,\,\overline{\langle\K_{y}\rangle}_{\textrm{I}}=0,
\eeq
implying that neither $\hat{\mathcal{C}}_y$ nor $\hat{\mathcal{K}}_y$ store relevant information within this phase. 

As it is clearly seen in Fig. \ref{fig:paneldeg}(c), the behavior of phase III is the exact opposite. This implies that only $\hat{\mathcal{C}}_y$ and $\hat{\mathcal{K}}_y$ are constants in this phase, and that
\beq
\overline{\langle \C_{x}\rangle}_{\textrm{III}}=0,\,\,\,\overline{\langle\K_{x}\rangle}_{\textrm{III}}=0,\,\,\,\overline{\langle\C_{y}\rangle}_{\textrm{III}}=2\sum_{n}\textrm{Re}\left(c_{n,+}c_{n,-}^{*}\bra{E_{n,-}}\C_{y}\ket{E_{n,+}}\right),\,\,\,\overline{\langle\K_{y}\rangle}_{\textrm{III}}=2\sum_{n}\textrm{Im}\left(c_{n,+}c_{n,-}^{*}\bra{E_{n,-}}\K_{y}\ket{E_{n,+}}\right),
\eeq
implying that the expectation values of $\hat{\mathcal{C}}_x$ and $\hat{\mathcal{K}}_x$ are irrelevant within this phase.

Summarizing, in order to characterize the dynamics of each phase, we need the following sets of constants of the motion. For phase I, the set is $\{\C_{x},\K_{x},\hat{\Pi}\}$. For phase II, $\{\hat{\Pi}\}$. And for phase III, $\{\hat{\Pi},\C_{y},\K_{y}\}$. 

\subsection{Extended generalized microcanonical ensemble}\label{GME}

Here, we provide a description of the order parameters in terms of a statistical ensemble. In \cite{Corps2022arXiv,Corps2022PRB}, a generalization of the standard microcanonical ensemble was devised to predict these order parameters in terms of constants of motion characterizing the excited-state quantum phases. The statistical ensemble proposed in these works deals with the case of a single ESQPT separating the spectrum into two distinct phases.  Coming back to the previous discussion on the aLMG, we note that neither $\C_{x}$ and $\hat{\mathcal{K}}_x$, nor $\C_{y}$ and $\hat{\mathcal{K}}_y$, can
be used to define a statistical ensemble because they only commute with the projectors into the energy subspaces corresponding to the first ESQPT or above the second ESQPT, respectively, but not with the entire Hamiltonian. This issue may be fixed by defining the closely related operators
\beq\label{eq:xtildes}
\widetilde{\mathcal{C}}_{x}\equiv \mathbb{I}_{\eps<\eps_{c_{1}}}\C_{x}\mathbb{I}_{\eps<\eps_{c_{1}}}, \,\,\,\widetilde{\mathcal{K}}_{x}\equiv \mathbb{I}_{\eps<\eps_{c_{1}}}\K_{x}\mathbb{I}_{\eps<\eps_{c_{1}}},
\eeq
and
\beq\label{eq:ytildes}
\widetilde{\mathcal{C}}_{y}\equiv \mathbb{I}_{\eps>\eps_{c_{2}}}\C_{y}\mathbb{I}_{\eps>\eps_{c_{2}}}, \,\,\,\widetilde{\mathcal{K}}_{y}\equiv \mathbb{I}_{\eps>\eps_{c_{2}}}\K_{y}\mathbb{I}_{\eps>\eps_{c_{2}}},
\eeq
where $\mathbb{I}_{\eps<\eps_{c_{1}}}\equiv \sum_{n}\theta_{n}\hat{P}_{n}$, $\hat{P}_{n}$ is the projector onto the eigenstates with energy $\eps_{n}$ and $\theta_{n}=1$ if $\eps_{n}<\eps_{c_{1}}$ and $\theta_{n}=0$ if $\eps>\eps_{c_{1}}$, and equivalently $\mathbb{I}_{\eps>\eps_{c_{2}}}\equiv \sum_{n}\gamma_{n}\hat{P}_{n}$ with $\gamma_{n}=1$ if $\eps_{n}>\eps_{c_{2}}$ and $\gamma_{n}=0$ if $\eps<\eps_{c_{2}}$. 

The expectation values $\langle \widetilde{\mathcal{C}}_{x,y}\rangle$ and $\langle \widetilde{\mathcal{K}}_{x,y}\rangle$  are identical to $\langle \C_{x,y}\rangle$ and $\langle \K_{x,y}\rangle$ if $\eps<\eps_{c_{1}}$ (for $x$) and $\eps>\eps_{c_{2}}$ (for $y$), but they vanish outside these intervals. The advantage of the operators in Eqs. \eqref{eq:xtildes} and \eqref{eq:ytildes} is that they commute with the Hamiltonian, and therefore, they may be used to build the statistical ensemble of our interest. Building on the ideas of \cite{Corps2022arXiv,Corps2022PRB}, we propose the following form for the density matrix of the (extended) generalized microcanonical ensemble (GME):
\beq\label{eq:rhogme}
\hat{\rho}_{\textrm{GME}}(E)=\hat{\rho}_{\textrm{ME}}(E)\left(\mathbb{I}+p\hat{\Pi}+c_{x}\widetilde{\mathcal{C}}_{x}+k_{x}\widetilde{\mathcal{K}}_{x}+c_{y}\widetilde{\mathcal{C}}_{y}+k_{y}\widetilde{\mathcal{K}}_{y}\right),
\eeq
where 
\beq\label{eq:micro}
\hat{\rho}_{\textrm{ME}}(E)=\frac{1}{2(N_{\textrm{I}}+N_{\textrm{II}}+N_{\textrm{III}})}\sum_{n}\left(\ket{E_{n,+}}\bra{E_{n,+}}+\ket{E_{n,-}}\bra{E_{n,-}}\right)
\eeq
denotes the standard microcanonical ensemble \cite{Alessio2016} with equal probability of population of parity doublets $\ket{E_{n,+}}$ and $\ket{E_{n,-}}$ within a small window, $\Delta E$, around the average energy value, $\langle E\rangle=\textrm{Tr}[\hat{\rho}\hat{\mathcal{H}}]$.  The denominator in Eq. \eqref{eq:micro} is required to guarantee that $\textrm{Tr} [\hat{\rho}_{\textrm{GME}}]=1$; it accounts for the total number of energy levels, $N=N_I + N_{II} + N_{III}$, within the energy window $\Delta E$. We write it in terms of the number of parity doublets in each of the phases to facilitate the calculation of $\langle \widetilde{\mathcal{C}}_{x,y} \rangle$ and $\langle \widetilde{\mathcal{K}}_{x,y} \rangle$. Additionally, Eq. \eqref{eq:rhogme} shows an explicit dependence on five parameters, $p,c_{x,y},k_{x,y}\in\mathbb{R}$ which are to be fixed by the condition that $\Tr[\hat{\rho}_{\textrm{GME}}\hat{\Pi}]=\langle \hat{\Pi}\rangle$, $\textrm{Tr}[\hat{\rho}_{\textrm{GME}}\widetilde{\mathcal{C}}_{x,y}]=\langle \widetilde{\mathcal{C}}_{x,y}\rangle$, $\textrm{Tr}[\hat{\rho}_{\textrm{GME}}\widetilde{\mathcal{K}}_{x,y}]=\langle \widetilde{\mathcal{K}}_{x,y}\rangle$. These conditions may be cast as follows: 
\beq\label{eq:condp}
\langle \hat{\Pi}\rangle=p,
\eeq
\beq\label{eq:condx}
\langle \widetilde{\mathcal{C}}_{x,y}\rangle=c_{x,y}\frac{N_{\textrm{q}}}{N_{\textrm{I}}+N_{\textrm{II}}+N_{\textrm{III}}}, \,\,\,
\langle \widetilde{\mathcal{K}}_{x,y}\rangle=k_{x,y}\frac{N_{\textrm{q}}}{N_{\textrm{I}}+N_{\textrm{II}}+N_{\textrm{III}}},
\eeq
with $q=\textrm{I}$ for the $x$ operators and $q=\textrm{III}$ for the $y$ operators. 

This ensemble reproduces a number of properties related to symmetry-breaking at the core of DPTs-I: 

(i) $\hat{\rho}_{\textrm{GME}}$ has off-diagonal elements in the Hamiltonian 
eigenbasis only if $c_{x}\neq 0$ and/or $k_{x}\neq 0$, or if $c_{y}\neq 0$ and/or $k_{y}\neq 0$. The first case is relevant when $\eps<\eps_{c_{1}}$, and thus the parity-breaking observable $\hat{J}_{x}$ may give rise to non-vanishing values, $\textrm{Tr}[\hat{\rho}_{\textrm{GME}}\hat{J}_{x}]\neq0$. The second case is important when $\eps>\eps_{c_{2}}$ instead, and it is the second component of the spin operator, $\hat{J}_{y}$, which may give rise to non-vanishing values, $\textrm{Tr}[\hat{\rho}_{\textrm{GME}}\hat{J}_{y}]\neq0$. 

For a state with energy $E_{n,k}\leq E_{c_{1}}$, in the energy eigenbasis $\{\ket{E_{n,+}},\ket{E_{n,-}}\}$ the GME can be written as the $2\times 2$ matrix 
\beq\label{eq:matrixbelow}
\rho_{n}(E_{n,k}\leq E_{c_{1}})=\frac{1}{2}\mqty(1+p && c_{x}-ik_{x} \\ c_{x}+ik_{x} && 1-p),
\eeq
whereas for a state $E_{n,k}\geq E_{c_{2}}$ this is 
\beq\label{eq:matrixabove}
\rho_{n}(E_{n,k}\geq E_{c_{2}})=\frac{1}{2}\mqty(1+p && k_{y}-ic_{y} \\ k_{y}+ic_{y} && 1-p).
\eeq
Note, however, that not every initial state leads to symmetry-broken equilibrium states if $\eps<\eps_{c_{1}}$ or $\eps>\eps_{c_{2}}$: indeed, if, due to the initial state chosen, $c_{x}=k_{x}=0$, it is still possible to have $\textrm{Tr}[\hat{\rho}_{\textrm{GME}}\hat{J}_{x}]=0$, and the same holds true for $c_{y}$, $k_{y}$ and $\hat{J}_{y}$. In other words, in the spectral phases $\eps<\eps_{c_{1}}$ and $\eps>\eps_{c_{2}}$ it is possible to find nonzero order parameters given by the long-time averages of $\hat{J}_{x}$ and $\hat{J}_{y}$, respectively. Each of these two spectral phases may be described by a \textit{different} order parameter.

(ii) $\hat{\rho}_{\textrm{GME}}$ is always diagonal in the parity eigenbasis in the intermediate spectral region $\eps_{c_{1}}<\eps<\eps_{c_{2}}$. In the energy eigenbasis $\{\ket{E_{n,+}},\ket{E_{n,-}}\}$, the GME is written, for a state with energy between both ESQPTs, as
\beq\label{eq:matrixint}
\rho_{n}(E_{c_{1}}<E_{n,k}<E_{c_{2}})=\frac{1}{2}\mqty(1+p && 0 \\ 0 && 1-p).
\eeq
Thus, if all states populated by a quench lie in this region, then all parity-breaking observables give vanishing values for any initial condition. 

Before we end this section, we address the mathematical form of the GME. The density matrix in Eq. \eqref{eq:rhogme} is a linear function of the observables commuting with the Hamiltonian. We should note that the form of the ensemble itself is not unique: for example, multiplicative matrices of the type of the generalized Gibbs ensemble, $\hat{\rho}\propto e^{-\beta\hatmath{H}-\sum_{q=x,y}\lambda_{q}^{c}\hatmath{C}-\sum_{q=x,y}\lambda_{q}^{k}\hatmath{K}}$, are also possible. Yet, the linear form of $\hat{\rho}_{\textrm{GME}}$ is the simplest it can take, which is why we have chosen it. Also, note that the width of the energy window, $\Delta E$, is a controllable parameter of the GME; therefore, its proposed form may be valid even if the quantum state is not tightly narrow in energy. 

Therefore, the ensemble Eq. \eqref{eq:rhogme} is well suited to describe order parameters caused by DPTs-I.

\section{Quench protocol} \label{protocol}

Dynamical phase transitions are commonly revealed through measurements of a wavefunction prepared in an initial state and then taken out of equilibrium. This is achieved by means of a quantum \textit{quench}. Usually, an initial state is prepared in a (initial) Hamiltonian obtained by fixing all control parameters of the model, in our case $\qham_{\textrm{ini}}\equiv \qham(\mathbf{\Theta}_{\textrm{ini}})$ where for notation convenience all control parameters are contained in $\mathbf{\Theta}_{\mu}\equiv (\xi_{\mu},\alpha_{\mu})$; we call\footnote{Throughout this paper, the total time elapsed since the initial state is initially chosen at $t=0$ will be denoted with a subscript, while the parameters of the Hamiltonian where the state evolves will be made explicit inside a parenthesis.} this state $\ket{\Psi_{t=0}(\mathbf{\Theta}_{\textrm{ini}})}$. Then the set of control parameters is changed abruptly to a final value, $\mathbf{\Theta}_{\textrm{ini}}\to\mathbf{\Theta}_{\textrm{fin}}$ that defines a final Hamiltonian, $\qham_{\textrm{fin}}\equiv \qham(\mathbf{\Theta}_{\textrm{fin}})$, where the system evolves. Measurements are then performed in the final Hamiltonian as the wavefunction evolves in time. 

Unfortunately, in our model it is not possible to perform a single quench between the two symmetry-breaking phases. The reason lies in the geometric interpretation that can be associated with any quench protocol. Let us imagine that we want to perform a quench between the low-energy and the high-energy symmetry-breaking phases in Fig. \ref{spectrum}, say $\xi_{\textrm{ini}} \rightarrow \xi_{\textrm{fin}}$ with $\xi_{\textrm{ini}} > \xi_{\textrm{fin}}$. Relying on the Helmann-Feynman theorem, it can be shown that this is only possible if the tangent line at a given eigenvalue with $\xi_{\textrm{ini}}$ crosses the two ESQPTs occurring between the two symmetry-breaking phases at two intermediate values $\xi_{1}$ and $\xi_{2}$, $\xi_{\textrm{ini}} > \xi_1 > \xi_2 > \xi_{\textrm{fin}}$ \cite{Perez2011}. But it is clearly seen in Fig. \ref{spectrum} that there is no energy level whose tangent line crosses the two ESQPTs in the interval $\xi \in [0,1]$. Therefore, we cannot lead the system from one symmetry-breaking phase to the other by means of a single quench. 

To solve this problem, in this paper, we go a step further with respect to the common protocol described above, letting the wavefunction evolve during a time of our choice in an \textit{intermediate} Hamiltonian, $\qham_{\textrm{int}}\equiv \qham(\mathbf{\Theta}_{\textrm{int}})$; that is, we perform a double quench \cite{Kennes2018,Cheraghi2023}. The protocol consists of the following steps:

(i) An initial state of the form
\beq\label{eq:initialstate}
\ket{\Psi_{0}(\mathbf{\Theta}_{\textrm{ini}})}=\sqrt{p}\ket{E_{\max,+}(\mathbf{\Theta}_{\textrm{ini}})}+e^{i\phi}\sqrt{1-p}\ket{E_{\max,-}(\mathbf{\Theta}_{\textrm{ini}})},
\eeq
is first prepared at $\mathbf{\Theta}_{\textrm{ini}}$. This state is a superposition of the two most excited states of $\qham_{\textrm{ini}}$ with different parities, i.e., $\qham_{\textrm{ini}}\ket{E_{\max,\pm}(\mathbf{\Theta}_{\textrm{ini}})}=E_{\max,\pm}(\mathbf{\Theta}_{\textrm{ini}})\ket{E_{\max,\pm}(\mathbf{\Theta}_{\textrm{ini}})}$ with $\hat{\Pi}\ket{E_{\max,\pm}(\mathbf{\Theta}_{\textrm{ini}})}=\pm \ket{E_{\max,\pm}(\mathbf{\Theta}_{\textrm{ini}})}$; only in the infinite-size limit, these states are degenerate, and \eqref{eq:initialstate} is stationary. Here, $p\in[0,1]$ is the probability of each eigenstate in the superposition, while $\phi\in[0,2\pi)$ is an arbitrary phase between them. Note that Eq. \eqref{eq:initialstate} is normalized for all values of $p$ and $\phi$, $\bra{\Psi_{0}(\mathbf{\Theta}_{\textrm{ini}})}\ket{\Psi_{0}(\mathbf{\Theta}_{\textrm{ini}})}=1$. 

(ii) At $t=0$, the initial value of the control parameters of the Hamiltonian is quenched to an intermediate value, $\mathbf{\Theta}_{\textrm{int}}$; thus, the initial state is taken out of equilibrium and dynamics begins. Notwithstanding, the key point is that {\em the system remains in a symmetry-breaking phase of the same nature that the one in the initial state}. The time evolution of the initial state in $\qham_{\textrm{int}}$ is given by the Schr\"{o}dinger equation ($\hbar=1$) generated by $\qham_{\textrm{int}}$,
\beq\label{eq:timeevoint}
\ket{\Psi_{t}(\mathbf{\Theta}_{\textrm{int}})}=e^{-i\qham_{\textrm{int}}t}\ket{\Psi_{0}(\mathbf{\Theta}_{\textrm{ini}})}=\sum_{k=\pm}\sum_{n}c_{n,k}^{\textrm{int}}e^{-iE_{n,k}(\mathbf{\Theta}_{\textrm{int}})t}\ket{E_{n,k}(\mathbf{\Theta}_{\textrm{int}})},
\eeq
where the coefficients $c_{n,k}^{\textrm{int}}\equiv \langle E_{n,k}(\mathbf{\Theta}_{\textrm{int}})|\Psi_{0}(\mathbf{\Theta}_{\textrm{ini}})\rangle$ are the overlap of the initial state with the eigenstates of the intermediate Hamiltonian. 

(iii) We let the wavefunction evolve up to a time $t=\tau_{\textrm{int}}$, which we will use as a control parameter. Even though the system remains in the same phase during all this time, we will see later that the value of $\tau_{\textrm{int}}$ determines the phase that the system reaches after the final quench.

(iv) Finally we perform a second quench leading the system to a final Hamiltonian $\mathbf{\Theta}_{\textrm{int}}\to\mathbf{\Theta}_{\textrm{fin}}$, where it is again left to evolve, 
\beq\label{eq:timeevofin}
\ket{\Psi_{\tau_{\textrm{int}}+t}(\mathbf{\Theta}_{\textrm{fin}})}=e^{-i\qham_{\textrm{fin}}t}\ket{\Psi_{\tau_{\textrm{int}}}(\mathbf{\Theta}_{\textrm{int}})}=\sum_{k=\pm}\sum_{n}c_{n,k}^{\textrm{fin}}e^{-iE_{n,k}(\mathbf{\Theta}_{\textrm{fin}})t}\ket{E_{n,k}(\mathbf{\Theta}_{\textrm{fin}})},
\eeq
where now $c_{n,k}^{\textrm{fin}}\equiv \langle E_{n,k}(\mathbf{\Theta}_{\textrm{fin}})|\Psi_{\tau_{\textrm{int}}}(\mathbf{\Theta}_{\textrm{int}})\rangle$ is the overlap of the intermediate wavefunction evaluated at $\tau_{\textrm{int}}$ with each of the eigenstates of the final Hamiltonian, $\qham_{\textrm{fin}}$. 

In this paper, we have made the following choices for the initial, intermediate, and final values of the Hamiltonian control parameters: $\mathbf{\Theta}_{\textrm{ini}}=(0.6,-2)$, $\mathbf{\Theta}_{\textrm{int}}=(0.2,-0.8)$, and $\mathbf{\Theta}_{\textrm{fin}}=(0.5,-0.6)$. A physical interpretation of the quench protocol can be gained by looking at the corresponding classical dynamics, which the quantum dynamics approach in the limit of large collective spin length $j\to\infty$. In Fig. \ref{fig:contours} we have represented such a classical analogy. We highlight the following facts:

(a) Figure \ref{fig:contours}(a) shows the initial state Eq. \eqref{eq:initialstate} with a green point. It lies in a disjoint region of the phase space in which all the orbits have $P<0$.

(b) The green orbit in Fig. \ref{fig:contours}(b) shows the time evolution after the first quench; it has been obtained by solving the corresponding Hamilton equations [see Eq. \eqref{eq:hamequations} below]. Its main feature is that $P(t)<0$ during all the times. This means that it is trapped in a disjoint region of the phase space with the same qualitative properties that the one that contains the initial state; this is why we say that the first quench keeps the system in the same phase. The markers in this figure show the position of the classical state at different values of the relaxation time: $\tau_{\textrm{int}}=0.5$ (blue square), $\tau_{\textrm{int}}=1.5$ (downward triangle) and $\tau_{\textrm{int}}=2.5$ (upward triangle). 

(c) Figure \ref{fig:contours}(c) shows the contour lines of the final Hamiltonian, together with the position of the system after the same values of the relaxation times as before, and the shape of the orbit that the system displays in the intermediate stage (with a dashed line). We can see that
if the state is left to evolve in the intermediate Hamiltonian only up to $\tau_{\textrm{int}}=0.5$ before the second quench is performed, then the final state is above all ESQPTs critical energies, $\varepsilon>\varepsilon_{c_{2}}$, and the system is trapped in a disjoint region of the phase space characterized by $P(t)<0$; that is, the system remains in the same phase than before the second quench. But if $\tau_{\textrm{int}}=1.5$, then $\varepsilon_{c_{1}}<\varepsilon<\varepsilon_{c_{2}}$, and the corresponding trajectory lies in a connected region, in which neither the sign of $P$ nor the sign of $Q$ remain constant; that is, the second quench implies a change of phase. And finally,
if $\tau_{\textrm{int}}=2.5$, then $\varepsilon<\varepsilon_{c_{1}}$, and the system gets trapped in a {\em different} disjoint region of the phase space, characterized by $Q<0$; that is, this quench implies a change of phase between two different disjoint regions. In general, as inferred from the dashed line in the figure, different relaxation times entail different topologies for the trajectories in the final stage. 

Note, however, that this interpretation is done in the classical limit. This means that it is expected to become more accurate as the system size is increased, but only if the initial state does not consist in a superposition of different phase-space regions. We will come back to this point in the next section.

\section{Dynamical phase transitions-I: order parameters}\label{DPT-I}

The first kind of dynamical phase transitions, DPTs-I, is related to the long-time average of the expectation value of a physically relevant observable and how this long-time average changes as a control parameter is varied. Generally, the analysis goes as follows: after a quench is performed, the expectation value of an observable $\hat{\mathcal{O}}$ is measured in the quenched wavefunction as it evolves in time. The long-time average is calculated and it is compared against different choices of the control parameters involved in the quench.

However, here we tweak this traditional procedure since our different quenches are not generated by changing the control parameters of the model each time, but by the time $\tau_{\textrm{int}}$ that the initial state spends in the intermediate Hamiltonian before it is changed to the final Hamiltonian. 
The advantage of using an intermediate Hamiltonian is that we have access to any region of the phase space of the system described by the final Hamiltonian, as discussed above. After letting the initial state evolve in the intermediate Hamiltonian for $\tau_{\textrm{int}}$ units of time, it is quenched to the final Hamiltonian, $\mathbf{\Theta}_{\textrm{int}}\to\mathbf{\Theta}_{\textrm{fin}}$, and the expectation value of an observable is measured
\beq\label{eq:expectation}
\langle \hat{\mathcal{O}}(t)\rangle\equiv \bra{\Psi_{\tau_{\textrm{int}}+t}(\mathbf{\Theta}_{\textrm{fin}})}\hat{\mathcal{O}}\ket{\Psi_{\tau_{\textrm{int}}+t}(\mathbf{\Theta}_{\textrm{fin}})},
\eeq
where $\ket{\Psi_{\tau_{\textrm{int}}+t}(\mathbf{\Theta}_{\textrm{fin}})}$ is given by Eq. \eqref{eq:timeevofin} and the time $t$ is measured in the final Hamiltonian. The long-time average of such time evolution for a given $\tau_{\textrm{int}}$ is then calculated as in Eq. \eqref{eq:longtime}. 

In this section, we study two kinds of initial states of the form Eq. \eqref{eq:initialstate}, with parameters: (S1) $p=1/2$, $\phi=3\pi/2$, and (S2) $p=1/3$, $\phi=3\pi/5$. State S1 is really a classical initial state as it is fully localized in the bottom energy well with $P<0$; therefore, $\langle \C_{y}\rangle=1$ and $\langle \C_{x}\rangle=0$. However, state S2 represents a quantum coherent superposition of both classical wells in phase III; it does not have a classical counterpart, and for this reason $|\langle \C_{y}\rangle|\neq 1$. 

\begin{figure}
\begin{centering}
\includegraphics[scale=0.8]{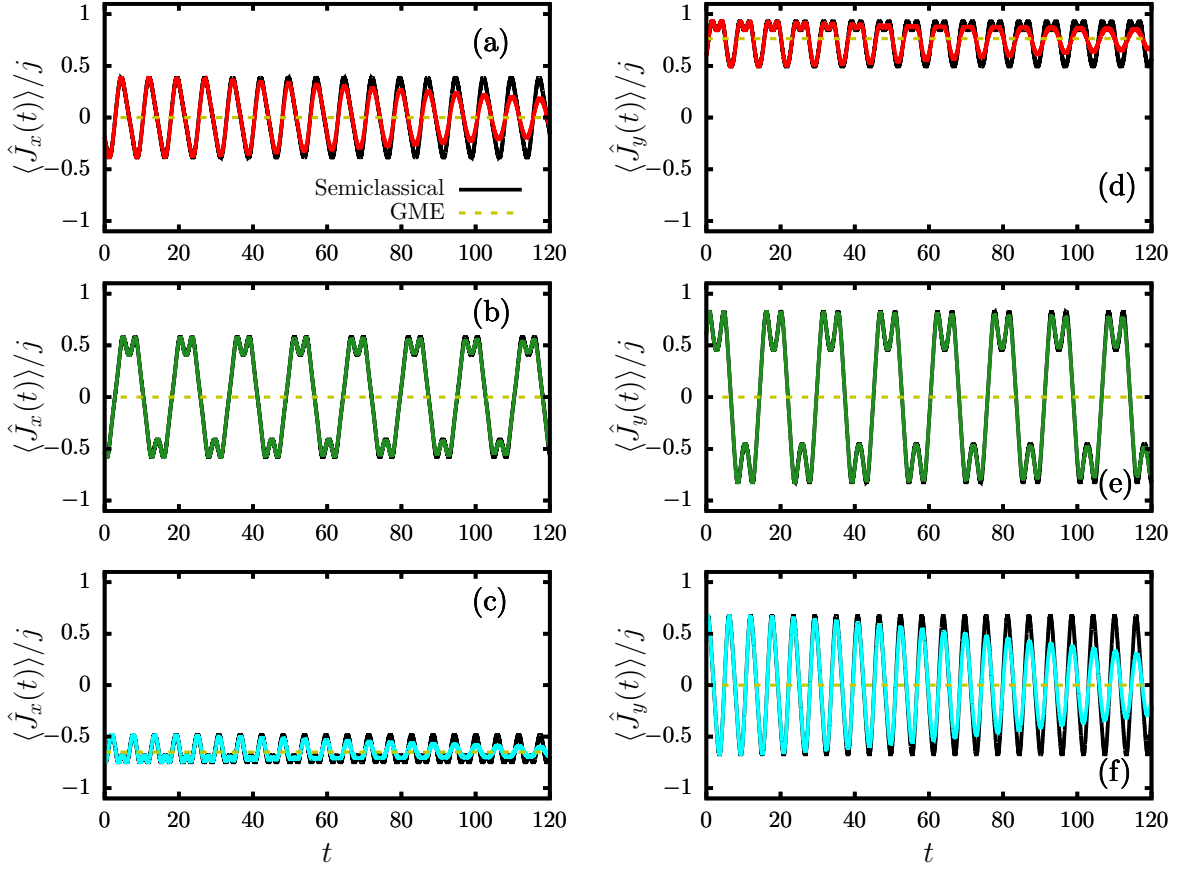}
\par\end{centering}
\caption{Signatures of DPTs-I. Time evolution of (a-c) $\hat{J}_{x}$ and (d-f) $\hat{J}_{y}$ as a function of time in the final Hamiltonian, $\mathbf{\Theta}_{\textrm{fin}}=(0.5,-0.6)$. In panels (a, d), the state evolves in the intermediate Hamiltonian during $t=0.5$ so the average energy of the quenched state is $\varepsilon<\varepsilon_{c_{1}}$; in panels (b, e), $t=1.5$ so $\varepsilon_{c_{1}}<\varepsilon<\varepsilon_{c_{2}}$; and in panels (c, f), $t=2.5$ so $\varepsilon<\varepsilon_{c_{2}}$. System size is $j=3200$, and the initial state \eqref{eq:initialstate} has $p=1/2$ and $\phi=3\pi/2$. The black lines represent the classical prediction for the time evolution, whereas the yellow dashed line shows the GME value for its infinite-time average. Color lines (red, green and cyan) represent the numerical calculations for each observable.}
\label{fig:evolution}
\end{figure}

We start with state S1. To illustrate the meaning of DPTs-I, we have represented the time expectation values of the first and second components of the collective spin operators, $\hat{J}_{x}$ and $\hat{J}_{y}$ together with the corresponding classical evolution and the results of the GME in Fig. \ref{fig:evolution}. The classical curves are obtained by solving the Hamilton equations spanned by the classical Hamiltonian Eq. \eqref{enfun}, 
\beq\label{eq:hamequations}
\frac{\textrm{d}Q}{\textrm{d}t}=\frac{\partial H}{\partial P}=P \left[(\alpha+2 \xi) Q^{2}+\alpha P^{2}-\xi+1\right],\,\,\frac{\textrm{d}P}{\textrm{d}t}=-\frac{\partial H}{\partial Q}=-Q \left[(\alpha+2 \xi) P^{2}+(\alpha+4 \xi) Q^{2}-5 \xi+1\right],
\eeq
by choosing as an initial condition $(Q(0),P(0))$ for the canonical coordinates corresponding to the state depicted in Fig. \ref{fig:contours}(a) (green dot), evolving it at $\mathbf{\Theta}_{\textrm{int}}$ until a time 
 $\tau_{\textrm{int}}$, $(Q(\tau_{\textrm{int}}),P(\tau_{\textrm{int}}))$. Then, we evolve this intermediate state at the final $\mathbf{\Theta}_{\textrm{fin}}$, and we obtain $(Q(\tau_{\textrm{int}}+t),P(\tau_{\textrm{int}}+t))$. The classical curves for $\hat{J}_{x}$ and $\hat{J}_{y}$ can be directly derived from Eq. \eqref{expected_values}.

In Fig. \ref{fig:evolution}, we observe a set of different patterns with oscillations around a given value, which is a defining feature of DPTs-I. The regularity of the patterns is due to the integrability of the model. Fig. \ref{fig:evolution}(a, d) show the case of $\tau_{\textrm{int}}=0.5$. The (finite-$j$) quantum results (red) match the classical curves (black) up to a certain time scale when the dynamics deviates from the classical counterpart. This behavior is common to collective spin models \cite{Corps2022arXiv,Corps2022PRB,Marino2022,Sciolla2011,Lerose2019}. As the system size increases, the quantum dynamics follows the classical results for longer times. It is clearly observed that while $\langle \hat{J}_{x}(t)\rangle$ oscillates around zero, $\langle \hat{J}_{y}(t)\rangle$ oscillates around a positive value. This behavior can be easily understood through the behavior of the classical limit of the model. For $\tau_{\textrm{int}}=0.5$, the state overlaps an orbit in the phase space of the final Hamiltonian characterized by a constant sign of $P<0$, since the trajectory is trapped in the lower (see Fig. \ref{fig:contours}(c)), high-energy classical well; however, the sign of $Q$ is not constant as the classical trajectory crosses the point $Q=0$. It follows from Eq. \eqref{expected_values} that, classically, $j_{x}\propto Q$ and $j_{y}\propto -P$, and therefore $\langle \hat{J}_{x}(t)\rangle$ can assume positive and negative values, with an average of zero, while $\langle \hat{J}_{y}(t)\rangle$ can only assume positive values.  A similar phenomenology is observed in the rest of the panels of Fig. \ref{fig:evolution}. Fig. \ref{fig:evolution}(b, e) show the case of $\tau_{\textrm{int}}=1.5$. Now, the state has an energy between both ESQPTs, $\varepsilon_{c_{1}}<\varepsilon<\varepsilon_{c_{2}}$, which is classically characterized by orbits that do not conserve either the sign of $Q$ or $P$ as the entire available phase space at that energy is accessible. Thus, both expectation values take positive and negative values, oscillating around a vanishing expectation value. Finally, Fig. \ref{fig:evolution}(c, f) depicts a situation opposite to that shown in Fig. \ref{fig:evolution}(a, d); indeed, in this case, the quenched state has energy $\varepsilon<\varepsilon_{c_{1}}$ and classically the orbit is trapped in the leftmost low-energy well, i.e., it is the sign of $Q<0$ that is conserved while that of $P$ is not.

The results shown in Fig. \ref{fig:evolution} neatly illustrate the defining signature of DPTs-I: a change in the long-time average of physical observables as a controllable parameter that drives the system through different energy regions is varied. In this case, the parameter inducing DPTs-I is the time spent by the initial state in the intermediate Hamiltonian, $\tau_{\textrm{int}}$, before quenching it to the final Hamiltonian.

\begin{figure}[h!]
\begin{centering}
\includegraphics[scale=0.8]{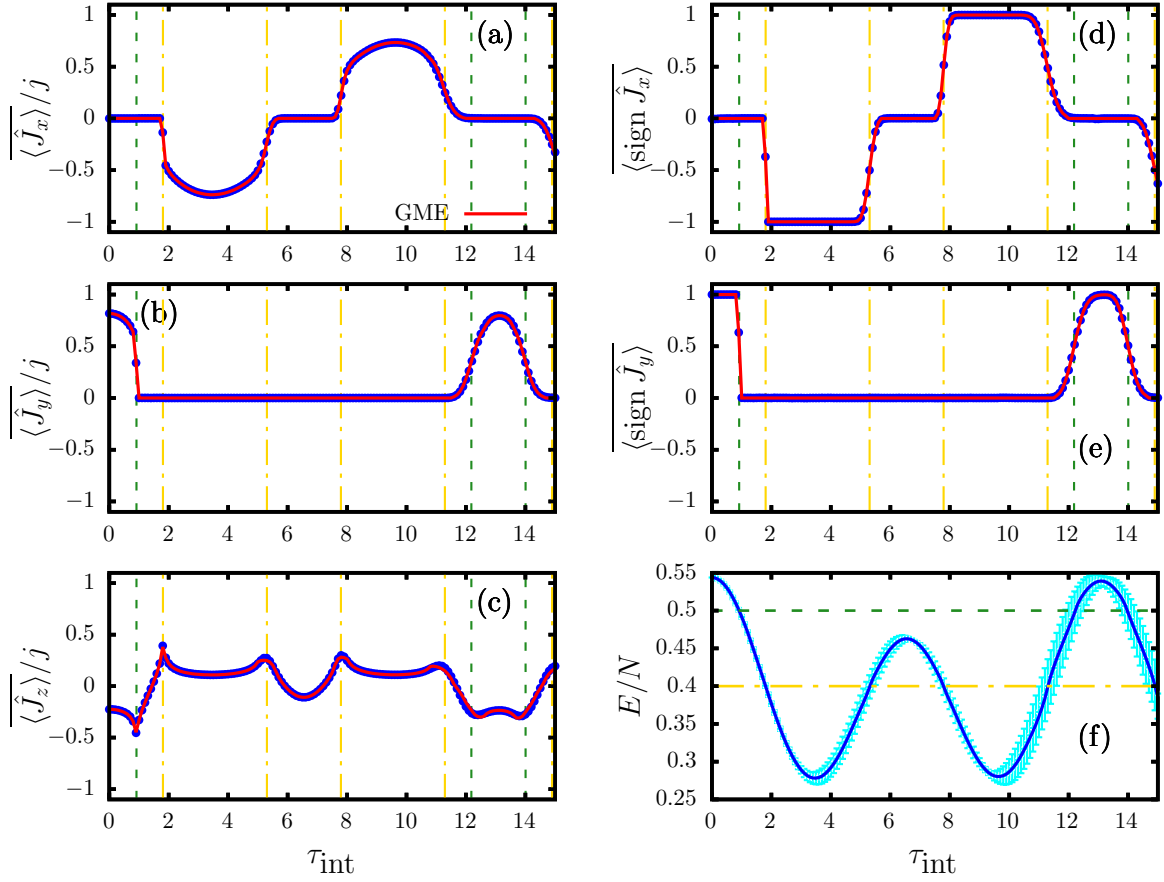}
\par\end{centering}
\caption{Further signatures of DPTs-I. (a-e) Long-time average of relevant observables and (f) average energy of the quench as a function of the time spent by the initial state in the intermediate Hamiltonian. The initial state is \eqref{eq:initialstate} with $p=1/2$, $\phi=3\pi /2$. The cyan error bars in (f) represent the standard deviation from the mean energy of the quenched state, $\sigma_{E}=(\sum_{n}\sum_{k=\pm}|c_{n,k}^{\textrm{fin}}|^{2}[\varepsilon_{n,k}(\mathbf{\Theta}_{\textrm{fin}})-\langle \varepsilon\rangle])^{1/2}$. The relaxation time in the final Hamiltonian is $\tau_{\textrm{fin}}=2000$. The evolution in the intermediate Hamiltonian is $\tau_{\textrm{int}}$, and the long-time average is taken in the final Hamiltonian over the time interval $t\in[0,\tau_{\textrm{fin}}]$. System size is $j=3200$. For the GME, the width of the energy window around the average energy of the initial state in the final Hamiltonian is $2\sigma_{E}$. The green dashed lines mark the ESQPT at $\varepsilon_{c_{2}}=\xi_{\textrm{fin}}=0.5$, while the yellow dashed lines mark the ESQPT at $\varepsilon_{c_{1}}=1+\alpha_{\textrm{fin}}=0.4$.  }
\label{fig:dpti}
\end{figure}

In Fig. \ref{fig:dpti}, we depict the average values of some observables as a function of $\tau_{\textrm{int}}$. In particular, starting from the initial state Eq. \eqref{eq:initialstate} in the initial Hamiltonian $\qham_{\textrm{ini}}$, we quench the system to the intermediate Hamiltonian $\qham_{\textrm{int}}$ and let the system evolve under this Hamiltonian during a time $\tau_{\textrm{int}}$, following Eq. \eqref{eq:timeevoint}. Exactly at $t=\tau_{\textrm{int}}$ the system is quenched again to the final Hamiltonian $\qham_{\textrm{fin}}$, and it is left to evolve for a time $\tau_{\textrm{fin}}=2000$. For each fixed value of $\tau_{\textrm{int}}$, the time expectation value, Eq. \eqref{eq:expectation}, of different observables is measured in the final Hamiltonian, and then its long-time average, Eq. \eqref{eq:longtime}, is computed. This is repeated for many quenches differing only in $\tau_{\textrm{int}}$.

Figure \ref{fig:dpti}(f) clearly shows that changing $\tau_{\textrm{int}}$ drives the system through the different energy regions of the final Hamiltonian, crossing several times the different ESQPTs. The passing of the state through these energy phases has an immediate consequence in the long-time averages of the spin operators in Fig. \ref{fig:dpti}(a-c) and the corresponding sign operators in Fig. \ref{fig:dpti}(d-e). These results may be interpreted in the same way as before.
For example, for $\tau_{\textrm{int}}\ll 1$, the quenched state has an average energy above all ESQPTs, $\eps>\eps_{c_{2}}$, and due to the particular parameters of the initial state Eq. \eqref{eq:initialstate} ($p=1/2$, $\phi=3\pi/2$), it becomes classically trapped within the bottom high-energy well. For this reason, $\langle \hat{J}_{x}(t)\rangle$ oscillates around zero, which is its long-time average, $\overline{\langle \hat{J}_{x}\rangle}=0$. On the same grounds, $\langle \hat{J}_{y}(t)\rangle$ oscillates around a \textit{positive} value, so $\overline{\langle \hat{J}_{y}\rangle}>0$.

This behavior is also reflected in the long-time averages of the signs $\hat{J}_{x}$ and $\hat{J}_{y}$, which become zero in the first case and unity in the second case. As $\tau_{\textrm{int}}$ increases, the quenched state enters the intermediate region $\eps_{c_{1}}<\eps<\eps_{c_{2}}$, where only $\hat{\Pi}$ is conserved, and classically corresponds to a well that covers both signs of $Q$ and $P$. For this reason, a DPT-I takes place, and $\overline{\langle \hat{J}_{y}\rangle}$ abruptly vanishes.

As $\tau_{\textrm{int}}$ increases further, the quenched state ends up in the spectral phase below all ESQPTs, $\eps<\eps_{c_{1}}$, and it gets trapped within the left low-energy classical well (see Fig. \ref{fig:contours}(c)), implying the constancy of $\C_{x}$ (sign$(\hat{J}_{x})$) and a negative value of the long-time average of $\hat{J}_{x}$, $\overline{\langle \hat{J}_{x}\rangle}<0$; here, a second DPT-I has occurred, but this time the relevant order parameter is given by $\hat{J}_{x}$, and not by $\hat{J}_{y}$.

Concerning $\hat{J}_{z}$, we can observe in Fig. \ref{fig:dpti}(c) that this operator is \textit{not} a parity-breaking observable because it can be diagonalized in the parity basis,  $[\hat{\Pi},\hat{J}_{z}]=0$. The various forms of non-analytic behavior observed in $\overline{\hat{J}_{z}}$ are not DPTs-I as $\hat{J}_{z}$ cannot be used as an order parameter; rather, they are due to the same phenomenon that causes the DPTs-I, the ESQPTs. In systems with a single classical freedom, $f=1$ [see Eq. \eqref{enfun}], the non-analyticities in the level density are directly transferred onto the expectation values of observables \cite{Caprio2008,Stransky2014,Cejnar2021}. Thus, the different peaks observed in $\overline{\langle \hat{J}_{z}\rangle }$ are a manifestation of the crossing of the quenched state from one spectral phase to another.

The rest of the results of Fig. \ref{fig:dpti}, for longer $\tau_{\textrm{int}}$, are explained following the same reasoning.  The only difference is that as $\tau_{\textrm{int}}$ increases,  an increasing diffusion of the quantum wave packet occurs, meaning that the quantum evolution deviates from the classical behavior and that the non-analytic points become smoother. This is shown with error bars Fig. \ref{fig:dpti}(f), accounting for the energy width of the corresponding wavefunction. 

Finally, we have also calculated the predictions of the extended GME, Eq. \eqref{eq:rhogme}, for the long-time averages discussed above. For this, we calculate the local density of states,
\beq\label{eq:ldos}
P(\eps)=\sum_{k=\pm}\sum_{n}|c_{n,k}^{\textrm{fin}}|^{2}\delta[\eps-\eps_{n,k}(\mathbf{\Theta}_{\textrm{fin}})],
\eeq
as well as its average energy $\langle \eps\rangle=\sum_{k=\pm }\sum_{n}|c_{n,k}^{\textrm{fin}}|^{2}\eps_{n,k}(\mathbf{\Theta}_{\textrm{fin}})$, and its width, $\sigma^{2}_{\eps}=\sum_{k=\pm }\sum_{n}|c_{n,k}^{\textrm{fin}}|^{2}[\eps_{n,k}(\mathbf{\Theta}_{\textrm{fin}})-\langle\eps\rangle]^{2}$. As in the standard microcanonical ensemble, to construct Eq. \eqref{eq:rhogme} we assume that the quench \textit{equally} populates all states in the final Hamiltonian within an energy window around the average energy, $[\langle \eps\rangle-\Delta\eps,\langle \eps\rangle+\Delta\eps]$. The width $\Delta\eps$ is an arbitrary quantity provided that $|\Delta\eps/\langle \eps\rangle| $ is sufficiently small but the number of levels within the window is large enough so that a statistical analysis is justified \cite{Alessio2016}. Here, we fix $\Delta \eps=2\sigma_{\eps}$. We then count the number of states within the energy window in phases I, II and III ($N_{\textrm{I}}$, $N_{\textrm{II}}$, $N_{\textrm{III}}$) to compute the parameters in Eqs. \eqref{eq:condp} and \eqref{eq:condx}, write the full extended GME density matrix $\hat{\rho}_{\textrm{GME}}$, and compare the numerical long-time averages $\overline{\langle \hat{\mathcal{O}}\rangle}$ with the GME value $\langle\hat{\mathcal{O}}\rangle_{\textrm{GME}}\equiv \textrm{Tr}\,[\hat{\rho}_{\textrm{GME}}\hat{\mathcal{O}}]$ 
 \cite{Corps2022arXiv,Corps2022PRB}. The agreement between the numerical results and the GME results is excellent in all cases, even for quite large values of $\tau_{\textrm{int}}$. For illustration purposes, the extended GME has been also represented in previous Fig. \ref{fig:evolution} (horizontal dashed line).

\begin{figure}[h!]
\begin{centering}
\includegraphics[scale=0.8]{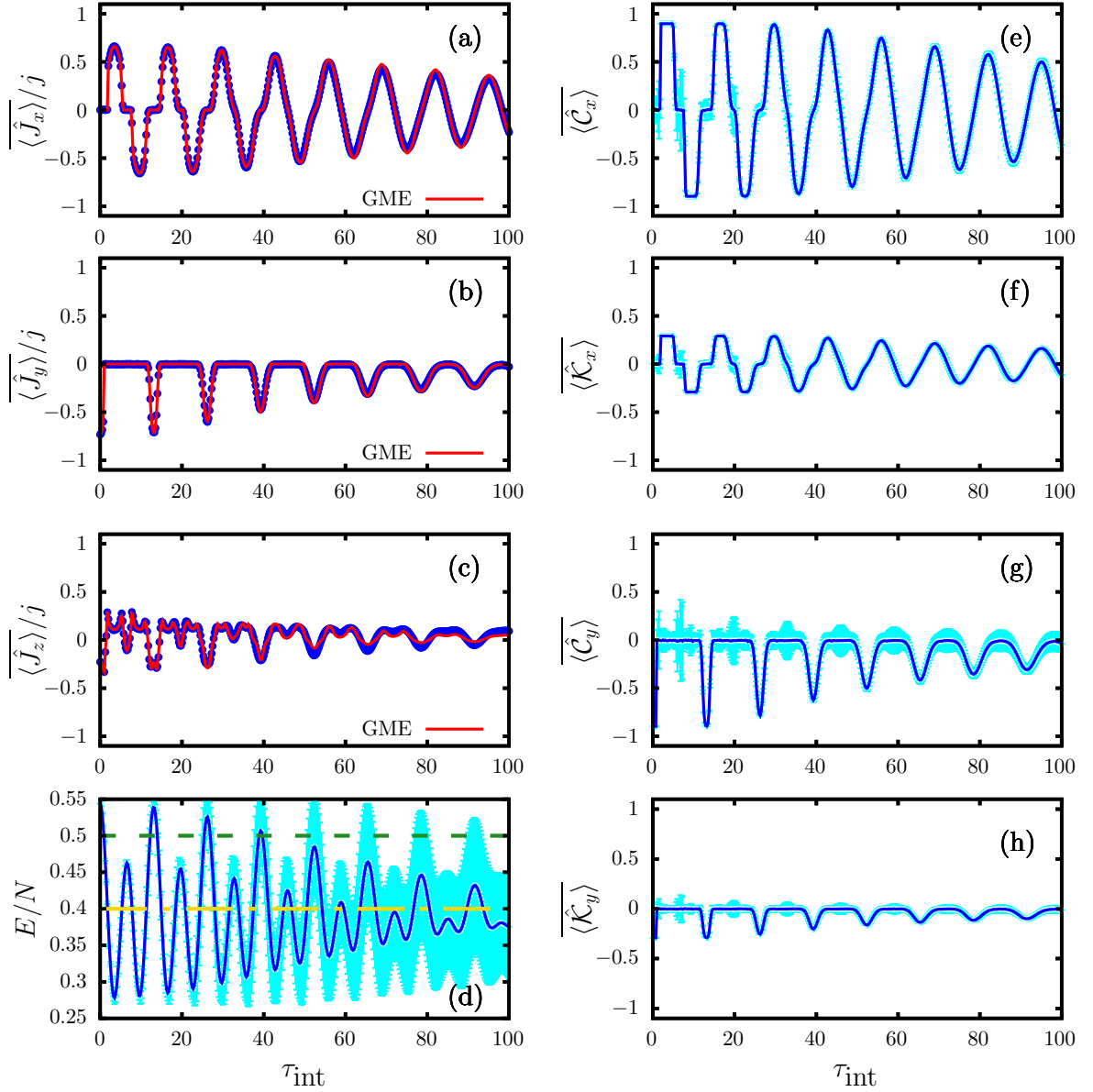}
\par\end{centering}
\caption{ The initial state is \eqref{eq:initialstate} with $p=1/3$, $\phi=3\pi /5$. (a-c) Long-time average of the angular momentum operators (blue dots) together with the GME predictions (red line). For the GME, the width of the energy window around the average energy of the initial state in the final Hamiltonian is $2\sigma_{E}$. The error bars in (d)-(h) (cyan) represent the standard deviation from the mean energy of the quenched state as well as the deviation from the long-time average of the corresponding observables. Hamiltonian is $\tau_{\textrm{fin}}=5000$. The evolution in the intermediate Hamiltonian is $\tau_{\textrm{int}}$, and the long-time average is taken in the final Hamiltonian over the time interval $t\in[0,\tau_{\textrm{fin}}]$. System size is $j=3200$. The green dashed lines mark the ESQPT at $\varepsilon_{c_{2}}=\xi_{\textrm{fin}}=0.5$, while the yellow dashed lines mark the ESQPT at $\varepsilon_{c_{1}}=1+\alpha_{\textrm{fin}}=0.4$.  }
\label{fig:dptiext}
\end{figure}

Now, we proceed with state S2. First of all, it is worth noting that the analogue of Fig. \ref{fig:evolution} makes no sense in this case, since the initial quantum state is a superposition of different classical phase-space regions, and therefore the resulting expectation values cannot be reproduced by a single classical trajectory, not even in the thermodynamic limit. Results are shown in Fig. \ref{fig:dptiext}, where we have extended the span of $\tau_{\textrm{int}}$ with respect to Fig. \ref{fig:dpti}. The effect of wavepacket diffusion is clearly illustrated in Fig. \ref{fig:dptiext}(d), where the error bars represent the width $\sigma_{E}$. For fast protocols with small $\tau_{\textrm{int}}$ the state is very localized in the final Hamiltonian eigenbasis. Notwithstanding, as the state consists in a superposition of different classical phase-space regions, both $\left| \langle \hat{\mathcal{C}}_y \rangle \right|$ and $\left| \langle \hat{\mathcal{K}}_y \rangle \right|$ are between 0 and 1. Furthermore, as $\tau_{\textrm{int}}$ increases the wavefunction becomes more and more delocalized. It is worth noting that for sufficiently long $\tau_{\textrm{int}}$ (for example $\tau_{\textrm{int}}\approx 80$), the wavepacket is so wide that it populates all three phases I, II and III simultaneously. As a consequence, all of the sign operators shown in Fig. \ref{fig:dptiext}(e-h) are non-zero at these times, but none of them is constant. Although the GME predictions for $\hat{J}_{x}$ and $\hat{J}_{y}$ in Fig. \ref{fig:dptiext}(a-b) match the numerical results at all times, the error increases significantly for $\hat{J}_{z}$ for longer $\tau_{\textrm{int}}$, as shown in Fig. \ref{fig:dptiext}(c). This results from the wavepacket diffusion that makes the state very delocalized in the energy eigenbasis, and also from the fact that the information of parity-breaking observables ($\hat{J}_{x,y}$) is directly encoded in the GME through the coefficients $c_{x,y}$ and $k_{x,y}$; no such information appears for observables that do not break parity ($\hat{J}_{z}$). Still, considering how ill-defined the average energy becomes at long $\tau_{\textrm{int}}$, this is to be expected. 

In summary, in the case of collective systems of this kind, DPTs-I order parameters are caused by the symmetry changes induced by the ESQPTs. Parity-breaking observables, such as $\hat{J}_{x}$ and $\hat{J}_{y}$, may give rise to order parameters via its long-time dynamics, these values being described by the (extended) GME Eq. \eqref{eq:rhogme}. As shown here, depending on the particular features of the classical limit, different symmetry-broken equilibrium states are possible. These define thermodynamically distinct phases in the energy spectrum, potentially described by different order parameters.

\section{Dynamical phase transitions-II: cusps in return probabilities}\label{DPT-II}

The second type of dynamical phase transitions, DPTs-II, denotes a form of non-analytic behavior exhibited by certain return probabilities of the initial wavefunction with its time-evolved self \cite{Heyl2018}. Indeed, it was proposed that the return amplitude of a time-evolved state (normally after a quench), $G(t)=\bra{\Psi_{0}(\lambda_{i})}e^{-i\qham(\lambda_{f})t}\ket{\Psi_{0}(\lambda_{i})}$, where $\lambda$ is a control parameter, may become non-analytic at certain critical times. This is usually identified by means of the rate function,
\beq
\label{eq:rate2}
\widetilde{r}_N \equiv - \frac{1}{N} \ln \textrm{SP} (t),
\eeq
which displays non-analytical kinks in the thermodynamic limit at the critical times. In this equation, $\textrm{SP}(t) = |G(t)|^2$. 

For systems with $\mathbb{Z}_{2}$ broken-symmetry phases, such as our model, a different measure was proposed in \cite{Heyl2014}, 
\beq\label{eq:pprp}
\mathcal{L}(t)\equiv \mathcal{L}_{+}(t)+\mathcal{L}_{-}(t)= |\bra{E_{0,+}(\lambda_{i})}e^{-i\qham(\lambda_{f})t}\ket{\Psi_{0}(\lambda_{i})}|^{2}+|\bra{E_{0,-}(\lambda_{i})}e^{-i\qham(\lambda_{f})t}\ket{\Psi_{0}(\lambda_{i})}|^{2},
\eeq
where $\ket{\Psi_{0}(\lambda_{i})}$ is some initial state and the indices $\pm$ denote the quantum numbers of some discrete $\mathbb{Z}_{2}$ symmetry. Commonly, $\ket{\Psi_{0}(\lambda_{i})}$ is taken to be a superposition of the symmetry-broken ground-state of an initial Hamiltonian $\qham(\lambda_{i})$, such as Eq. \eqref{eq:initialstate}, which is then taken out of equilibrium by a quantum quench. When the $\mathbb{Z}_{2}$ symmetry labeling the eigenstates of the Hamiltonian is parity, $\hat{\Pi}$, we call Eq. \eqref{eq:pprp} \textit{parity-projected return probability} (PPRP) \cite{Corps2022arXiv,Corps2022PRB}. It gives rise to a different rate function,
\beq
\label{eq:rate}
r_N \equiv - \frac{1}{N} \ln \mathcal{L} (t),
\eeq

In \cite{Corps2022arXiv,Corps2022PRB} a link between the behavior of these magnitudes and the presence of constants of motion like $\hat{\mathcal{C}}_{x,y}$ and $\hat{\mathcal{K}}_{x,y}$ has been proved. This link implies: (i) $\mathcal{L}(t) = \textrm{SP} (t)$ in any symmetry-breaking phase in which there are degenerate parity doublets; (ii) the main mechanism proposed for the presence on non-analytic kinks in $r_N(t)$ \cite{Heyl2014} does not work in the presence of degenerate parity doublets. Furthermore, it was numerically shown that an \textit{anomalous} DPT-II occurs in such a phase, in which the first non-analytic time occurs only after the first local maximum in the rate functions \cite{Homrighausen2017,Corps2022PRB}, and the finite-size scaling of $r_N(t)$ does not show a clear trend. Notwithstanding, this observation comes from quenches after which the system remains in the same symmetry-breaking phase. Hence, taking into account that it was shown in \cite{Corps2023PRB} that the ESQPT is not directly related with the generation of anomalous dynamical phases, it is interesting to see what happens when a quench between different symmetry-breaking phases is performed.

\begin{figure}
\begin{centering}
\includegraphics[scale=0.8]{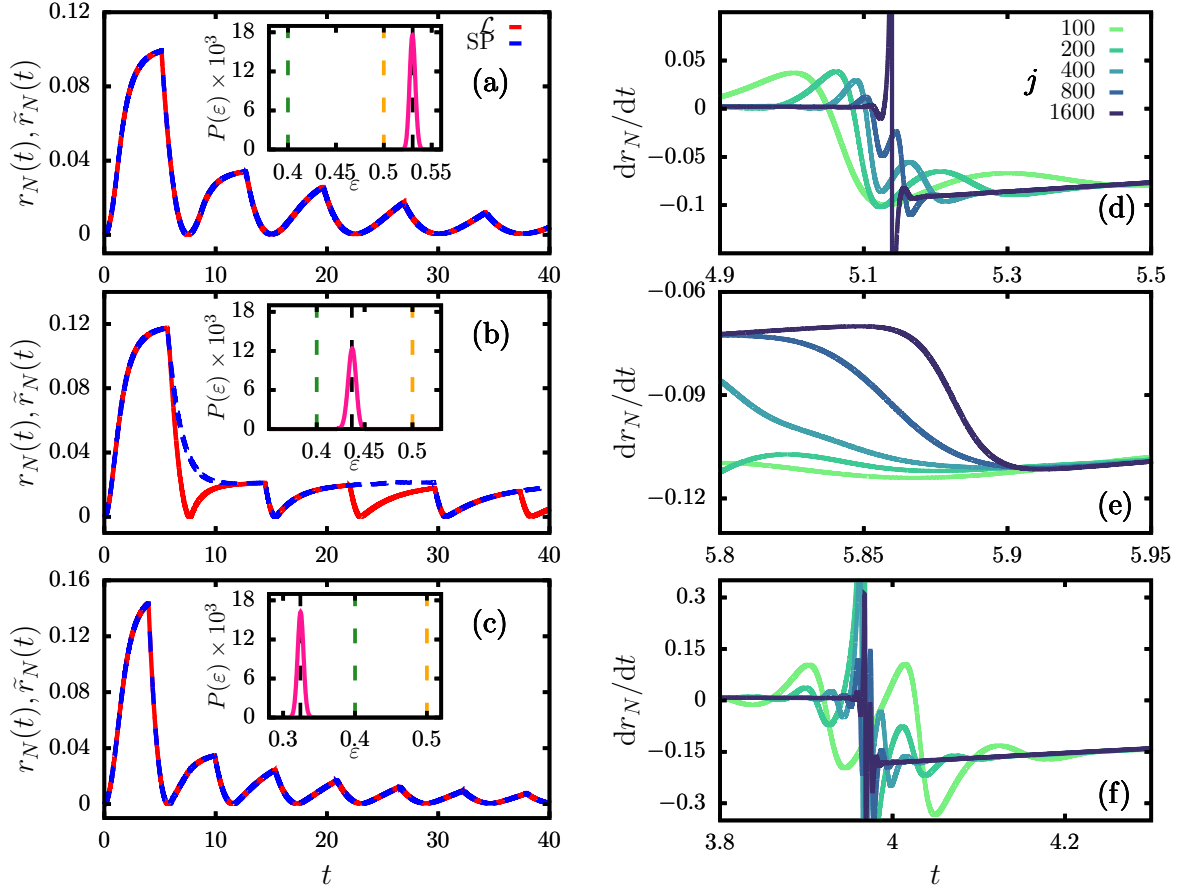}
\par\end{centering}
\caption{Signatures of DPTs-II. (a-c) Rate functions associated to the PPRP (full red line), $\mathcal{L}$, and to the survival probability (dashed blue line), $\textrm{SP}$, for a system size $j=1600$. The initial state \eqref{eq:initialstate} has $p=1/2$ and $\phi=3\pi/2$. This state evolves in the intermediate Hamiltonian during a time (a) $\tau_{\textrm{int}}=0.5$, (b) $\tau_{\textrm{int}}=1.5$ and (c) $\tau_{\textrm{int}}=2.5$ before the second quench is performed. The insets show the local density of states $P(\varepsilon)$ for the quenched state corresponding to the return probabilities in each main panel. Green and orange vertical dashed lines mark the ESQPTs critical energies $\varepsilon_{c_{1}}=1+\alpha_{\textrm{fin}}=0.4$ and $\varepsilon_{c_{2}}=\xi_{\textrm{fin}}=0.5$, while the black vertical dashed line represents the classical energy of the quenched state in the final Hamiltonian. (d-f) First derivative of the rate function of the PPRP of panels (a-c), respectively, for different system sizes $j$ [see legend in (d)].}
\label{fig:paneldptii1}
\end{figure}

We again prepare an initial state of the form Eq. \eqref{eq:initialstate}, with $p=1/2$ and $\phi=3\pi/2$, at an initial value of the set of control parameters, $\mathbf{\Theta}_{\textrm{ini}}$. We then quench the system to the intermediate Hamiltonian,  $\mathbf{\Theta}_{\textrm{ini}}\to\mathbf{\Theta}_{\textrm{int}}$, and let the state evolve according to Eq. \eqref{eq:timeevoint}. At time $t=\tau_{\textrm{int}}$, a final quench is performed, $\mathbf{\Theta}_{\textrm{int}}\to\mathbf{\Theta}_{\textrm{fin}}$, the evolution being given by Eq. \eqref{eq:timeevofin}. As after the first quench the system remains in the same symmetry-breaking phase than in the initial state, our aim is to study the possible DPTs-II emerging from the second quench. Thus, we study the PPRP and survival probability of the state evolving in the final Hamiltonian, Eq. \eqref{eq:timeevofin}, considering that its initial state is the state in the intermediate Hamiltonian at the last time before it was quenched to the final Hamiltonian, Eq. \eqref{eq:timeevoint} with $t=\tau_{\textrm{int}}$.  That is, we study the following version of the PPRP,

\beq\label{eq:pprp2}
\mathcal{L}^{\textrm{fin}}(t)\equiv \mathcal{L}^{\textrm{fin}}_{+}(t)+\mathcal{L}^{\textrm{fin}}_{-}(t)= |\bra{\Psi_{\tau_{\textrm{int}},+}(\mathbf{\Theta}_{\textrm{int}})}e^{-i\qham(\lambda(\mathbf{\Theta}_{\textrm{fin}}))t}\ket{\Psi_{\tau_{\textrm{int}}}(\mathbf{\Theta}_{\textrm{int}})}|^{2}+|\bra{\Psi_{\tau_{\textrm{int}},-}(\mathbf{\Theta}_{\textrm{int}})}e^{-i\qham(\lambda(\mathbf{\Theta}_{\textrm{fin}}))t}\ket{\Psi_{\tau_{\textrm{int}}}(\mathbf{\Theta}_{\textrm{int}})}|^{2},
\eeq
and the following version of the survival probability,
\beq\label{eq:sp}
\textrm{SP}^{\textrm{fin}}(t)=|\bra{\Psi_{\tau_{\textrm{int}}}(\mathbf{\Theta}_{\textrm{int}})}e^{-i\qham(\lambda(\mathbf{\Theta}_{\textrm{fin}}))t}\ket{\Psi_{\tau_{\textrm{int}}}(\mathbf{\Theta}_{\textrm{int}})}|^{2}.
\eeq
Here, $\ket{\Psi_{\tau_{\textrm{int}},\pm}(\mathbf{\Theta}_{\textrm{int}})}$ denotes the part of the time evolution of the initial state in the intermediate Hamiltonian with a definite value of parity, $\pm$, so that the full state evolving in the intermediate Hamiltonian is simply $\ket{\Psi_{\tau_{\textrm{int}}}(\mathbf{\Theta}_{\textrm{int}})}=\ket{\Psi_{\tau_{\textrm{int}},+}(\mathbf{\Theta}_{\textrm{int}})}+\ket{\Psi_{\tau_{\textrm{int}},-}(\mathbf{\Theta}_{\textrm{int}})}$, see Eq. \eqref{eq:timeevoint}. This separation is always possible because, $\hat{\Pi}$ being a conserved quantity, the evolution of the positive and negative parity sectors are completely decoupled in the Schr\"{o}dinger equation.

Fig. \ref{fig:paneldptii1} shows the results of the corresponding rate functions, $r_{N}(t)$ and $\widetilde{r}_{N}(t)$, Eqs. \eqref{eq:rate2} and \eqref{eq:rate}, for $j=1600$, in panels (a-c), and first derivatives of the rate function of the PPRP for different system sizes $j$, $\textrm{d}r_{N}/\textrm{d}t$ and $\textrm{d}\widetilde{r}_{N}/\textrm{d}t$, in panels (d-f). The insets of Figs. \ref{fig:paneldptii1}(a-c) show the distribution of populated states for the quenched state in the final Hamiltonian, Eq. \eqref{eq:ldos}, together with the corresponding classical expectation of the average energy. Rows differ by the value of $\tau_{\textrm{int}}$. 

Fig. \ref{fig:paneldptii1}(a) shows the rate functions for $\tau_{\textrm{int}}=0.5$, with a final population of states in phase III, $\eps>\eps_{c_{2}}$. We can see that in this symmetry-broken phase, $\mathcal{L}(t)=\textrm{SP}(t)$, exactly as predicted by the analytical results of \cite{Corps2022arXiv,Corps2022PRB}. The derivative of the rate function of $\mathcal{L}(t)$ is shown in Fig. \ref{fig:paneldptii1}(d): Even though the largest case, $j=1600$, is the less smooth, there is no clear tendency as a function of $j$; for example, the curve of $j=800$ is smoother than that of $j=400$. In this region, therefore, the behavior of the cusps in the thermodynamic limit seems unclear. This last fact is compatible with the results in \cite{Corps2022arXiv,Corps2022PRB}. But there is a remarkable difference between those results and the ones shown in \ref{fig:paneldptii1}(a). In this case, the first kink is observed in the first maximum of $r_N(t)$, whereas in \cite{Corps2022arXiv,Corps2022PRB} and in other works dealing with anomalous DPT-II \cite{Homrighausen2017} the first kink is always observed in the second maximum of DPT-II. Therefore, it is not clear whether this DPT-II can be called anomalous or not.

When the quenched state ends in the intermediate energy region $\eps_{c_{1}}<\eps<\eps_{c_{2}}$, Fig. \ref{fig:paneldptii1}(b) shows that $\mathcal{L}(t)$ and $\textrm{SP}(t)$ are not the same function, again as predicted by the results in \cite{Corps2022arXiv,Corps2022PRB}. This is because there are no level degeneracies in phase II. Yet, there are times when $\mathcal{L}(t)$ and $\textrm{SP}(t)$ either separate or become equal. The first time when this happens, around $t\approx 5.9$, is analyzed in Fig. \ref{fig:paneldptii1}(e). The first derivative of $r(t)$ is quite smooth for small system sizes (note, for example, the flatness of the case with $j=100$), but clearly approaches a step function as $j$ increases. That is, results in Fig. \ref{fig:paneldptii1}(e) seem compatible with a jump from $\sim -0.07$ to $\sim -0.11$ at $t\sim 5.9$ in the thermodynamic limit. These results are totally compatible with those of \cite{Corps2022arXiv,Corps2022PRB}.

Finally, when the quenched state ends below all ESQPTs, $\eps<\eps_{c_{1}}$, in phase I, Fig. \ref{fig:paneldptii1}(c) shows that we again have the equality $\mathcal{L}(t)=\textrm{SP}(t)$, with derivatives of a similar nature to that in Fig. \ref{fig:paneldptii1}(d) which do not clearly approach a clear discontinuity; in this case, the largest oscillations occur for $j=200,400$. Thus, the results from quenches between two different symmetry-breaking phases are qualitatively similar to the obtained from quenches between the same symmetry-breaking phase. In particular, we also observe in Fig. \ref{fig:paneldptii1}(c) that the first kink appears at the first maximum of $r_N(t)$.

\begin{figure}
\begin{centering}
\includegraphics[scale=0.8]{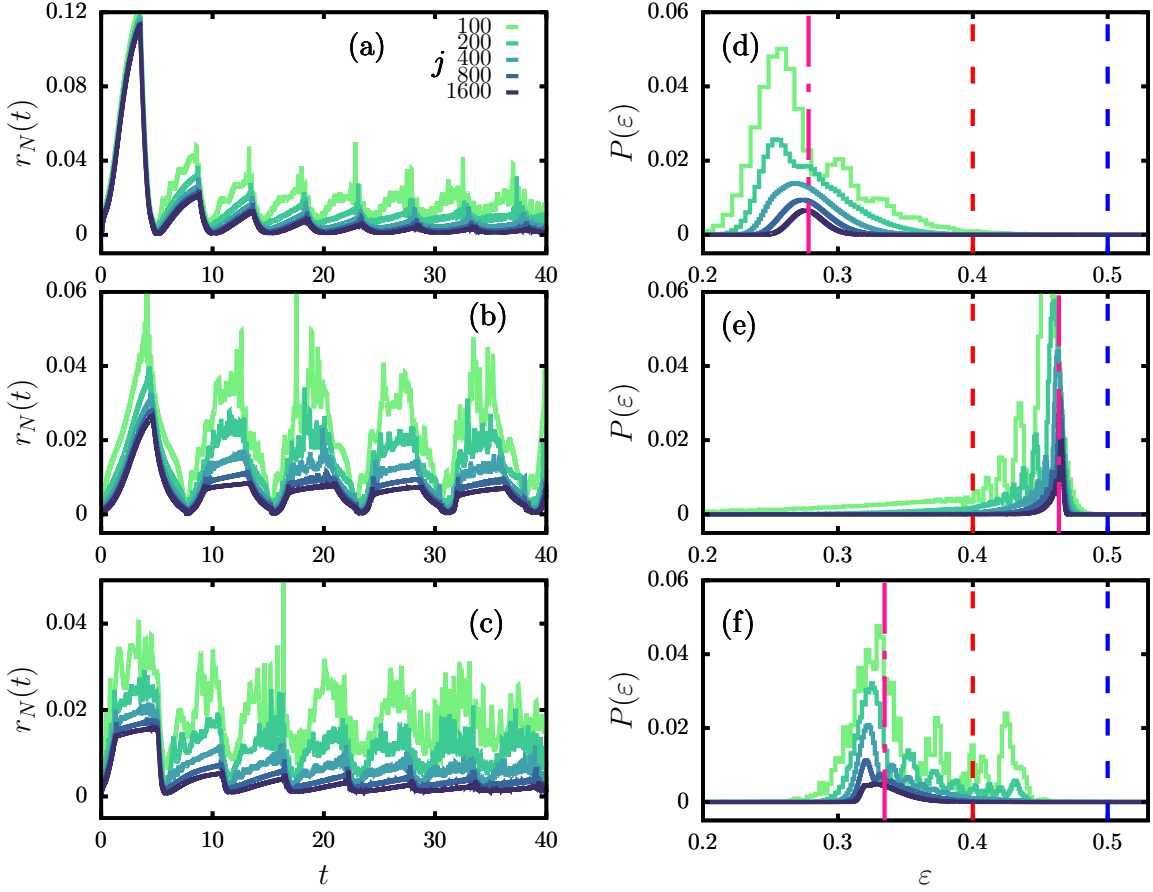}
\par\end{centering}
\caption{Further signatures of DPTs-II. (a-c) Rate functions associated to the PPRP, $\mathcal{L}$ for several system sizes $j$ [see legend in (a)]. The initial state is \eqref{eq:initialstate} with $p=1/2$ and $\phi=3\pi/2$. This state evolves in the intermediate Hamiltonian during a time (a) $\tau_{\textrm{int}}=3.5$, (b) $\tau_{\textrm{int}}=6.5$ and (c) $\tau_{\textrm{int}}=8.5$. (d-f) Local density of states $P(\varepsilon)$ for the quenched states corresponding to the return probabilities in (a-c), respectively. Red and blue vertical dashed lines mark the ESQPTs critical energies $\varepsilon_{c_{1}}=1+\alpha_{\textrm{fin}}=0.4$ and $\varepsilon_{c_{2}}=\xi_{\textrm{fin}}=0.5$, while the pink dashed-dotted line represents the classical average energy $\langle \varepsilon \rangle$.}
\label{fig:paneldptii2}
\end{figure}

Fig. \ref{fig:paneldptii2} depicts the rate function associated to the PPRP, $r_{N}(t)$, for several system sizes $j$ and for intermediate times longer than those considered in Fig. \ref{fig:paneldptii1}. The larger the system size, the sharper $r_{N}(t)$ in Figs. \ref{fig:paneldptii2}(a-c). However, the time spent by the state in the intermediate Hamiltonian is equally important. Therefore, in order to study how this magnitude scales with the size of the system, it is necessary to take into account not just the size of the system itself, but the time elapsed in the intermediate Hamiltonian.

Figure \ref{fig:paneldptii2}(a),(c) correspond to quenches between two different symmetry-breaking phases, whereas in Fig. \ref{fig:paneldptii2}(b) the final state is in a disordered phase (II).  This can be seen in Fig. \ref{fig:paneldptii2}(d-f), where the local density of states, Eq. \eqref{eq:ldos}, is shown, together with the corresponding classical expectation of the average energy.  The energies of the first and second ESQPTs, $\eps_{c_{1}}$ and $\eps_{c_{2}}$, are also indicated with vertical dashed lines. The average energy of the wavepacket is in phase I in panels (d) and (f), while it falls in phase II in (e), $\eps_{c_{1}}<\langle \eps \rangle<\eps_{c_{2}}$. For larger system sizes, the wavepacket becomes more localized as the quantum state approaches classical behavior, where it is fully localized. A remarkable result is that the shape of the first peak in $r_N(t)$ is different in the last case, showing a kind of plateau between two kinks. This may be due to the larger width of the wavefunction, which is already overlapping the disordered phase. In any case, results in Fig. \ref{fig:paneldptii2} show that DPTs-II, and especially the anomalous ones, constitute a very involved phenomenon. 

\section{Conclusions}\label{conclusions}

The usual scenario for the study of dynamical phase transitions involves a quantum system with two different phases, typically a symmetry-breaking phase and a disordered phase, and a quench leading the system from one phase to the other. In this work, we propose a protocol to explore a more complex situation, with three different phases: two {\em different} symmetry-breaking ones, and a disordered one. Its defining trademark is that it uses the relaxation time between two quenches as the only control parameter to explore the full dynamical phase diagram.

We apply this protocol to the anharmonic Lipkin-Meshkov-Glick model. The reason for this choice is that this model exhibits two excited-state quantum phase transitions, which split the spectrum into three different excited-state phases. Our first result is to show that there is a disordered phase, in which the parity is the only constant of motion (besides the Hamiltonian itself), and two symmetry-breaking phases, each one characterized by two {\em different} non-commuting constants of motion. Each of these two symmetry-breaking phases can be identified by two different order parameters: $\hat{J}_x$ and $\hat{J}_y$, respectively.

Then, we derive a generalized microcanonical ensemble depending on all the constants of motion, and we apply it to the study of dynamical phase transitions of type I. As main conclusions, we obtain: (i) there exist three different dynamical phases, in correspondence with the three different excited-state phases, and (ii) the generalized microcanonical ensemble provides a very good description of long-time averages, and a clear signature of the dynamical changes.

To deal with DPTs-II, we focus on the consequences of the second quench; the first one is devised to keep the system within the same phase that in the initial state. 
We have analyzed the appearance of non-analytic times in return probabilities, and the impact that the diffusion time of the initial state in the intermediate Hamiltonian may have in these dynamical features. We observe that DPTs-II are qualitatively different depending on whether or not there exists a second constant of motion after the second quench. The survival probability and the parity-projected survival probability are different only if the parity is the only constant of motion after the second quench; the times at which these two magnitudes depart from each other display finite-size precursors of a non-analytic behavior with a clear finite-size scaling. However, if the phase after the second quench has a second constant of motion besides parity, either the same that the phase in which the system is before this second quench or a different one, both survival probabilities are equal and the origin of the emerging non-analytic points is not clear. A second important observation is that this system does not display the typical anomalous DPTs-II when the quench leads the system onto a symmetry-breaking phase. On the one hand, the scaling of the rate functions is not clear, as usually observed in such cases. But, on the other hand, the first observed kink occurs in the first maximum of the rate function, and this is not typical for anomalous but for regular DPTs-II. As a final result, we show that all of these phenomena become blurred if the relaxation time in the intermediate stage is very long, due to the diffusion of the wave packet; and that more involved behaviors, like the presence of approximate plateaus between kinks in the rate function, also appear. 

\begin{acknowledgments}
 A. L. C., A. R. and P. P.-F. acknowledge financial support by the Spanish grants PGC-2018-094180-B-I00, PID2019-104002GB-C21, and PID2019-104002GB-C22  funded by Ministerio de Ciencia e Innovaci\'{o}n/Agencia Estatal de Investigaci\'{o}n MCIN/AEI/10.13039/501100011033 and FEDER "A Way of Making Europe". A. L. C. acknowledges financial support from `la Caixa' Foundation (ID 100010434) through the fellowship LCF/BQ/DR21/11880024. P. P.-F. is also supported by the Consejer\'{\i}a de Conocimiento, Investigaci\'on y Universidad, Junta de Andaluc\'{\i}a and European Regional Development Fund (ERDF) under project US-1380840 and it is also part of grant Groups FQM-160 and the project PAIDI 2020 with reference P20\_01247, funded by the Consejería de Economía, Conocimiento, Empresas y Universidad, Junta de Andalucía (Spain) and “ERDF—A Way of Making Europe”, by the “European Union” or by the “European Union NextGenerationEU/PRTR”.
\end{acknowledgments}

\end{document}